\documentclass{article}
\usepackage{caption,subcaption}
\usepackage{graphicx}
\usepackage{authblk}
\usepackage{pdfpages}
\usepackage{hyperref}
\usepackage{cleveref}

\PassOptionsToPackage{numbers, compress}{natbib}

\usepackage[final]{neurips_data_2024}

\usepackage[utf8]{inputenc} 
\usepackage[T1]{fontenc}    
\usepackage{hyperref}       
\usepackage{url}            
\usepackage{booktabs}       
\usepackage{amsfonts}       
\usepackage{nicefrac}       
\usepackage{microtype}      
\usepackage{xcolor}         

\title{The State of Data Curation at NeurIPS: An Assessment of Dataset Development Practices in the Datasets and Benchmarks Track}

\author[1,3]{Eshta Bhardwaj}
\author[2]{Harshit Gujral}
\author[2]{Siyi Wu}
\author[1]{Ciara Zogheib}
\author[1]{Tegan Maharaj}
\author[1,3]{Christoph Becker}
\affil[1]{Faculty of Information, University of Toronto}
\affil[2]{Department of Computer Science, University of Toronto}
\affil[3]{Digital Curation Institute, University of Toronto}

\begin{document}

\maketitle

\vspace{-4mm}
\begin{abstract}
\vspace{-2mm}
Data curation is a field with origins in librarianship and archives, whose scholarship and thinking on data issues go back centuries, if not millennia.
The field of machine learning is increasingly observing the importance of data curation to the advancement of both applications and fundamental understanding of machine learning models -- evidenced not least by the creation of the Datasets and Benchmarks track itself. 
This work provides an analysis of recent dataset development practices at NeurIPS through the lens of data curation. 
We present an evaluation framework for dataset documentation, consisting of a rubric and toolkit developed through a thorough literature review of data curation principles. 
We use the framework to systematically assess the strengths and weaknesses in current dataset development practices of 60 datasets published in the NeurIPS Datasets and Benchmarks track from 2021-2023. 
We summarize key findings and trends. Results indicate greater need for documentation about environmental footprint, ethical considerations, and data management. We suggest targeted strategies and resources to improve documentation in these areas and  provide recommendations for the NeurIPS peer-review process that prioritize rigorous data curation in ML.  We also provide guidelines for dataset developers on the use of our rubric as a standalone tool.
Finally, we provide results in the format of a dataset that showcases aspects of recommended data curation practices.  
Our rubric and results are of interest for improving data curation practices broadly in the field of ML as well as to data curation and science and technology studies scholars studying practices in ML. Our aim is to support continued improvement in interdisciplinary research on dataset practices, ultimately improving the reusability and reproducibility of new datasets and benchmarks, enabling standardized and informed human oversight, and strengthening the foundation of rigorous and responsible ML research.

\end{abstract}

\section{Introduction} \label{introduction}

The NeurIPS Datasets and Benchmarks (D\&B) track was created in 2021 to enhance the development of datasets in line with the exponential growth of applications of machine learning (ML). A key challenge this track aimed to address is that of datasets being used  outside their original scope as benchmarks \cite{koch_reduced_2021, paullada_data_2021, raji_ai_2021, yeung_announcing_2021} -- among other potential issues, this creates the possibility of field-level overfitting, as well as unanticipated ethical and privacy problems. NeurIPS sought to address this by encouraging the development of new datasets for ML, improving the quality of datasets being produced, and emphasizing the importance of the role of data within ML. The introduction of new, tailored peer-review guidelines also enabled and incentivized publication of datasets and benchmarks. The D\&B track is thus uniquely positioned to influence and guide the quality and ethicality of datasets being released and bolster responsible dataset development practices.

Recent research in fairness, accountability, and transparency has proposed to reduce bias in models, via datasets, through the improvement of dataset documentation \cite{gebru_datasheets_2021,li_data-centric_2022,liang_advances_2022,madaio_co-designing_2020,miceli_documenting_2021,muller_forgetting_2022,pushkarna_data_2022,sambasivan_everyone_2021}. Particularly, recent research on ethical data curation for ML datasets \cite{Leavy_Siapera_O’Sullivan_2021} emphasizes the adoption of concepts and processes from library and archival studies and digital curation in order to improve the documentation of dataset contents and data design decisions \cite{Leavy_Siapera_O’Sullivan_2021,Colavizza_Blanke_Jeurgens_Noordegraaf_2022,Jo_Gebru_2020,Thylstrup_2022}.\textbf{ Data curation}, a subset of digital curation and information science, is ``...the activity of managing data throughout its life cycle; appropriately maintaining its integrity and authenticity; ensuring that it is properly appraised, selected, securely stored, and made accessible; and supporting its usability in subsequent technology environments.'' \citep[p.~203]{noonan_data_2014}. We recently translated data curation concepts for the ML dataset development context \cite{bhardwaj_machine_2024}. The results enable us to apply data curation to the field of ML so that dataset development practices can be evaluated and improved.  

\textbf{Contributions.} 
Our goal is to document and improve the standard of dataset development in NeurIPS so that future benchmarks and datasets can be effectively found, easily accessed, ethically used, consistently evaluated, and appropriately reused. 
To these ends, we present a systematic dataset documentation evaluation framework to organize the assessment of curation practices for ML dataset development.  The framework is composed of a rubric and toolkit developed through an iterative multi-stage process to arrive at the rubric elements relevant for evaluation and corresponding criteria for their assessment, as well as supplementary material packaged as a toolkit to aid in the application of the rubric. Here, we establish the feasibility of this framework as an auditing tool for dataset documentation by performing a systematic evaluation of a sample set of 60 datasets published in the NeurIPS D\&B track between 2021-2023. We analyze the assessments to evaluate how data curation is currently performed in ML and show how it can be improved. 
Our results demonstrate that documentation quality varies widely across datasets and reveal a lack of documentation and reflexivity on environmental footprint, the situatedness and non-neutral nature of data, ethical considerations, and data management. We recommend how to improve this in Section \ref{discusion} and make a proposal for NeurIPS peer-review changes. 

\section{Background} \label{background}

\subsection{The Importance of Data in Machine Learning}

Increasingly, machine learning research has turned towards the improvement of data to improve model results and fundamental understanding. Research areas include the role of data (distribution) in learning theory and generalization \cite{memgen, pacbounds, datapac, fantastic, sloppy, overparam}, explicitly data-centric machine learning \cite{cfp_dlmr}, the construction of datasets \cite{almohaimeed_thos_2023,gomez_speech_2023,lamm_retail-786k_2024,peddi_put_2023,pingle_l3cube-mahasent-md_2023,shinde_windset_2024}, addressing a greater number of areas and problem domains \cite{arnaiz-rodriguez_towards_2024,deng_computational_2024,kohli_towards_2024,lee_beyond_2024,vysogorets_towards_2024,zhao_measuring_2024}, the development of ethical models/frameworks around AI and data \cite{cfp_aies} within sound and music computing, computer vision, natural language interfaces, and more \cite{barnett_ethical_2023,cherepanova_deep_2023,deshpande_responsible_2022,mcilroy-young_mimetic_2022,seymour_systematic_2023}. 

NeurIPS has responded to the rising urgency and recognized impact of data through the introduction of the Datasets and Benchmarks (D\&B) track \cite{yeung_announcing_2021}. Submissions to the NeurIPS D\&B track highlight aspects of data work critical to the development of machine learning datasets. Specifically, \cite{peng_mitigating_2021} looks at impacts of unmanaged citations of dataset derivatives, continued usage of datasets after their retraction, ambiguous dataset documentation, non-restrictive and ineffective licensing, and lack of long-term data stewardship. A data quality  framework developed for datathons \cite{mougan_how_2023}, is also applicable for the examination of data quality of ML datasets as it covers 5 broad quality dimensions. A checklist for ethical data collection of human-centric computer vision datasets highlights that  further emphasis should be placed on moving away from general-purpose datasets to clearly-defined dataset collection, providing recommendations for obtaining consent and maintaining privacy, and proposing how to examine, acknowledge, and enhance dataset diversity \cite{andrews_ethical_2023}. 

In 2023 NeurIPS released a Code of Ethics \cite{noauthor_neurips_nodate} to supplement the NeurIPS Code of Conduct \cite{noauthor_neural_nodate}. The code of ethics includes information on how to report and prevent harms from research that involves human participants, data concerns including privacy, consent, etc., and societal harms such as concerns of safety, security, discrimination, harassment, environment, human rights, and bias. Alongside this, there were ethics guidelines released for reviewers \cite{noauthor_2023_2023} which point reviewers to checklists and a framework for evaluating general ethical conduct, as well specifications for human-related data, data concerns around compliance, consent, and regulations, and negative societal impacts. Guides and further information about how to report on these areas have been released by NeurIPS and others \cite{carolyn_ashurst_guide_2020,noauthor_neurips_2021}. Since 2021, the D\&B track has seen immense growth and success, from 174 of 484 submitted papers accepted in 2021 to 322 of 987 submitted papers accepted in 2023 \cite{noauthor_neurips_2021-1,noauthor_neurips_2023}.


As the landscape of data-focussed ML research continues to evolve, there remain open challenges for the field to tackle. For example, a review of AI impact statements in NeurIPS papers from 2021 \cite{liu_examining_2022} finds that `agency' and `responsibility' are two key themes in the statements whereby dataset creators feel they are not in control of the negative impacts of their work if it was to be misused by malicious users. Authors typically reassign the responsibility to identify and safeguard against potential negative impacts to other practitioners, or state that the potential negative impacts of their work are the same as those that exist for that domain, i.e., further work is needed to recognize contributions as non-neutral and take accountability of how design decisions impact outcomes \cite{liu_examining_2022}.

\vspace{-2mm}
\subsection{Data Curation}

Data curation's influences as a field include information sciences, digital libraries, and archival sciences \cite{palmer_foundations_2013,ross_digital_2012,abrams_foundational_2015,lee_wheres_2011,chao_data_2014,Jo_Gebru_2020}. It thus has deep-rooted and established methods and discourse on how to maintain large amounts of data and manage ethical concerns. Data curation as a component of digital curation takes a \textbf{lifecycle approach} to the management of digital data \cite{noonan_data_2014}. Lifecycle approaches divide the digital curation practice into stages, which differ in their details but have in common a broad view including the pre-use, use, and post-use of a dataset. For example, the Digital Curation Center's model specifies `conceptualize', `create or receive', `appraise and select', `ingest', `preservation action', `store', `access, use, and reuse', and `transform' \cite{higgins_dcc_2008}. Each stage of curation consists of activities, designated roles and responsibilities, specified technical, legal, ethical, and operational considerations as well as policies that institutionalize the discussed activities and responsible parties \cite{higgins_dcc_2008}. 

 ML studies on data practices (``...what and how data are collected, managed, used, interpreted, reused, deposited, curated, and so on...'' \citep[p.~55]{Borgman_2015}) also called dataset development and data work most closely resemble examinations of the processes of data curation (despite often using the term to connote data collection, e.g., \cite{madaio_co-designing_2020,li_data-centric_2022,Holstein_Wortman_Vaughan_Daumé_Dudik_Wallach_2019}). Other works in this space discuss the importance of documentation and propose new frameworks for it \cite{bender_data_2018,Chmielinski_Newman_Taylor_Joseph_Thomas_Yurkofsky_Qiu_2022,gebru_datasheets_2021}, review intrinsic and extrinsic biases in the dataset development process \cite{muller_forgetting_2022,passi_problem_2019,Muller_Lange_Wang_Piorkowski_Tsay_Liao_Dugan_Erickson_2019}, and highlight the power dynamics involved in data quality, data work, and documentation \cite{Miceli_Posada_Yang_2022}.  

In recent work \cite{bhardwaj_machine_2024}, we performed a thorough literature review of data curation practices, translated them to be applied in a machine learning setting, and presented a preliminary analysis of a few NeurIPS D\&B track datasets, focussed on the interdisciplinary process of adopting data curation for ML. Particularly, we highlighted the need for tools to translate the standards for transparency and accountability. In contrast to previously mentioned studies, our framework enables the application of \textit{data curation} principles and concepts in practice.

Other authors also point to the need for dataset creators to \textit{actively} document and steward their datasets \cite{peng_mitigating_2021}, in contrast to much of the static documentation which is common in ML. These recommendations can be translated into practice-based processes: by seeing dataset development in ML as data curation and adopting its norms and practices, the NeurIPS community and ML research practices broadly can gain an elevated standard of documentation and resulting benefits to model performance, responsibility, and fundamental understanding. 

\section{Methods} \label{methods}

\textbf{Research Questions.} What are the strengths and weaknesses of NeurIPS dataset documentation practices considered through a data curation lens? In other words, how well curated are NeurIPS datasets and benchmarks? To study this, we examine (1) What constitutes a well curated dataset? (2) How feasible is the adoption of data curation principles to assess ML datasets? and (3) What is the state of data curation at NeurIPS and how can it be further advanced?

\textbf{Approach.} We developed an evaluation framework to assess data documentation practices, i.e., curation processes, and applied it to recently published ML datasets in the NeurIPS datasets and benchmarks track. This track was precisely chosen because of its relevance in publishing such contributions but also in influencing the quality of datasets that are accepted. The evaluation framework consists of a rubric used to evaluate dataset documentation and design decisions and a toolkit which supplements the rubric by providing additional information on how to apply the rubric effectively. This framework was applied to manually assess 60 ML datasets in three steps. 

\begin{enumerate}
    \item We established the \textbf{initial construction and design} of our evaluation framework consisting of the rubric and toolkit and 
    reviewed the \textbf{preliminary feasibility} of the framework by applying it to 25 datasets across 4 rounds \cite{bhardwaj_machine_2024}. In these rounds, we continued to develop the framework iteratively based on the evaluation results \cite{bhardwaj_machine_2024}. We reflected and reported on the initial process of designing the framework such as the benefits resulting from the diverse perspectives of an interdisciplinary team, the lessons learned while applying the framework, and how we used the data from the initial application of the rubric to iteratively refine it and yield more consistent evaluations \cite{bhardwaj_machine_2024}. 
    \item We \textbf{examined the consistency in application} by measuring inter-rater reliability (IRR). To claim that our framework is consistent, reliable, and accordingly feasible, we conducted another round of evaluations consisting of 5 datasets (round 4 and disagreement review) with the framework fully developed. We therefore address RQ 1 with our most updated version of the framework and RQ 2 with the final fourth round of evaluations that firmly establishes the framework’s reliability through iteratively improving IRR results.  
    \item We applied this framework to assess 30 additional datasets to \textbf{evaluate current practices} of data curation in ML dataset development and areas where improvement was needed (RQ 3).
\end{enumerate}


\textbf{Evaluation Framework.} We grounded our framework in data curation principles, emphasizing documentation, transparency, and ethical considerations. We started with key aspects of data curation relevant to ML and followed with iterative refinement through internal reviews and adjustments to evaluation criteria, guided by digital curation lifecycle models, FAIR data principles, and environmental sustainability and justice considerations. The rubric, provided in the Appendix, consists of 18 elements across five categories. In \cite{bhardwaj_machine_2024}, we presented results from a training round and rounds 1-3. After round 3, we updated and refined the criteria for 13 elements and added additional guidance in the toolkit for interpreting authenticity, reliability, and representativeness. We present the up-to-date version of the framework along with the changes made between versions and their rationale in the Appendix.

The \textbf{scope} category has 2 elements, `context, purpose, motivation' and `requirements', which emphasize the requirement for a dataset creation plan and addressing intrinsic biases. The \textbf{ethicality and reflexivity} category has 4 elements, `ethicality', `domain knowledge and data practices', `context awareness', and `environmental footprint', covering a range of documentation requirements to increase reflection and accountability in the dataset creation process. The \textbf{data pipeline} category includes `data collection', `data processing', and `data annotation', prompting reflection on how and why choices were made and their implications. The \textbf{data quality} category contains  `suitability', `representativeness', `authenticity', `reliability', and `structured documentation', to ensure the consideration of a broad set of qualities that impact how well a dataset can be appropriately and responsibly reused. The \textbf{data management} category covers FAIR principles \cite{wilkinson_fair_2016} - findability, accessibility, interoperability, and reusability -  included to evaluate the transparency of data management considerations. Each rubric element is assessed on minimum standard criteria (with a score of `pass' or `fail') that detail the expected level of documentation. Elements that pass the minimum standard are also assessed on a standard of excellence (with a score of `full', `partial', or `none'). Therefore, the conceptualization of the rubric defines what a well-curated dataset must document. The \textbf{toolkit} is a supplementary resource that introduces concepts from data curation and serves as a manual to the rubric. It contains instructions and guidance on how to evaluate datasets, how to interpret specific elements, guiding principles, recommendations, FAQ, sample evaluations, a glossary, and further readings. The toolkit is provided in the Appendix. 


\textbf{Iterations.} In order for data curation to provide robust norms for ML dataset development, our framework has to  \emph{enable consistent use}. To evaluate consistency, we  measured inter-rater reliability (IRR) as we iteratively refined the rubric over multiple rounds of evaluation. The preliminary stage of refining the rubric occurred across the first 4 rounds of evaluations (namely training, round 1, round 2, round 3) \cite{bhardwaj_machine_2024}. Each round involved improvements based on feedback and observations, ensuring the rubric and toolkit were effectively refined and validated. This was structured around several key activities. We began with a training round to help reviewers become acquainted with the rubric and foundational data curation concepts, significantly reducing initial discrepancies and increasing IRR in the upcoming rounds. Following each evaluation round, we gathered feedback from reviewers and identified specific areas of the rubric that needed adjustments to better convey the expectations and reduce ambiguity. We refined definitions, provided clearer examples, and better aligned the rubric elements with practical evaluation scenarios. This established preliminary feasibility. 

\textbf{Consistency.} To establish the framework's feasibility and consistency, we performed additional rounds of evaluations. Across training to round 4, three reviewers assessed each of the 30 datasets in a fully crossed design \cite{mcgraw_forming_1996} thus we calculated IRR using a two-way mixed, consistent, average-measures intra-class coefficient (ICC) to assess the consistency of the raters' evaluations of rubric elements measured on an ordinal scale across subjects \cite{hallgren_computing_2012}. In rounds 3 and 4, we additionally performed a ``disagreement review''. Once reviewers had completed their evaluations, they reviewed other evaluations and engaged in a brief discussion period in which they could debate, review, and update their scores and comments. Given the interpretative nature of the framework, this collaborative disagreement review enabled improved understanding of the rubric concepts and mitigated potential errors such as overlooking provided documentation. It also led to greater consistency while simultaneously leveraging the diverse perspectives of reviewers to enhance the richness and accuracy of the dataset evaluations.  
With the framework's feasibility established, we evaluated additional datasets. 

\textbf{Application.} To understand precisely how data curation can contribute to the advancement of ML dataset documentation practices, we performed a final round of evaluations (round 5). In this round, we evaluated 30 datasets published in the NeurIPS D\&B track from 2021-2023. A full list of evaluated datasets from all rounds can be found in the Appendix. The datasets were randomly selected after filtering all published papers at the NeurIPS D\&B track for dataset contributions. The filtering process is described in the Appendix. In the final round, four reviewers performed double coding for 30 datasets, each reviewing on average 15 datasets, including a disagreement review. Accordingly, we measured IRR with a one-way mixed, consistent, average-measures intra-class coefficient (ICC). After the disagreement review, additional corrections were made for consistency, see Appendix. All 60 dataset evaluations and analysis files can be found hosted on Zenodo \cite{Bhardwaj_2024_neuripsdataset}.


\textbf{Analysis.} We analyze to what extent criteria were fulfilled for 1) each dataset and 2) each rubric element. This enables a review of whether data curation can provide guidance for documentation for NeurIPS datasets and precisely in what capacity that guidance is needed.

\section{Results} \label{results}

\begin{figure*}[h!]
    \centering
    \begin{subfigure}[b]{0.4\textwidth}
        \includegraphics[width=\textwidth]{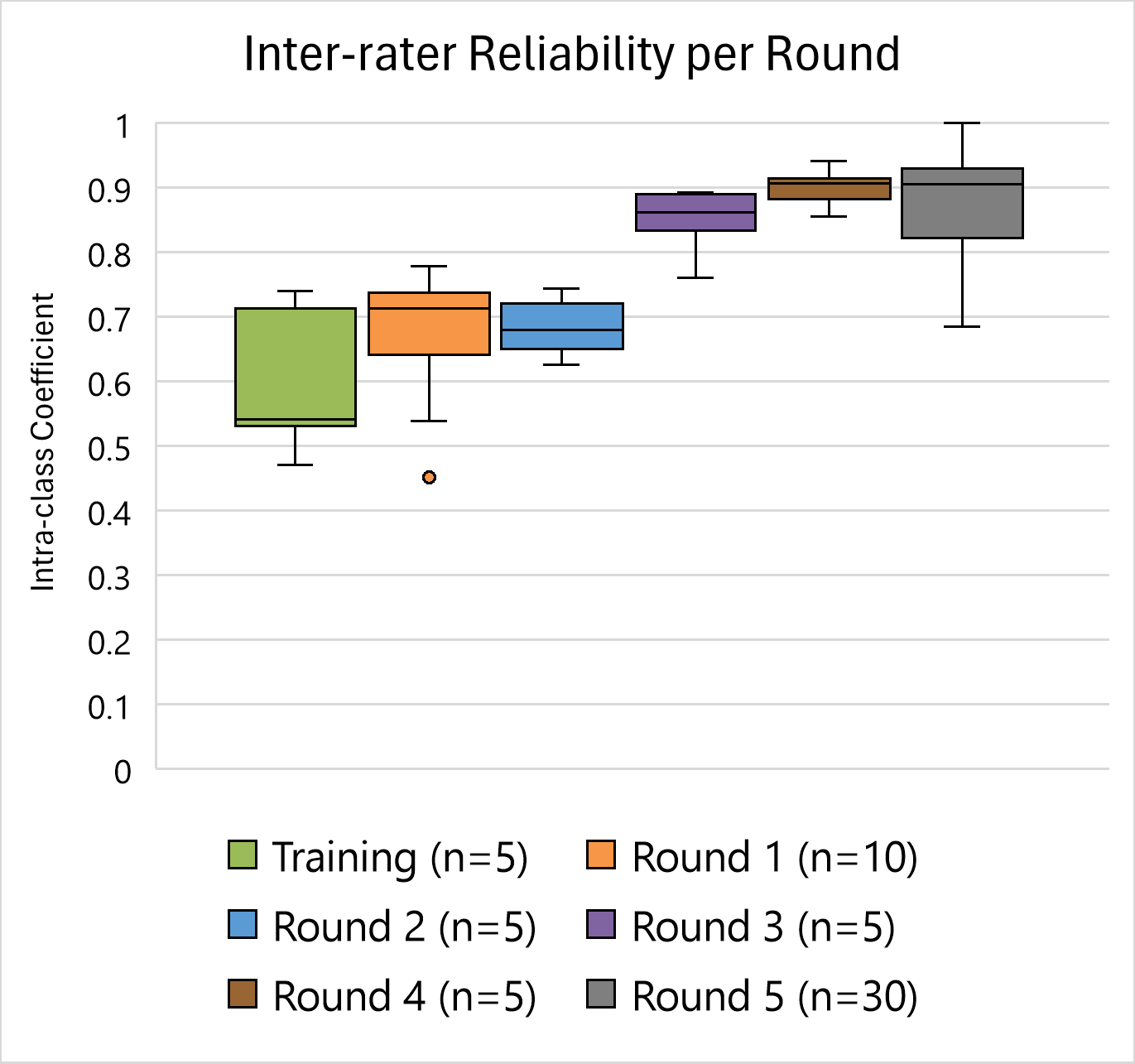}
        \caption{IRR for evaluation rounds}
        \label{fig:fig1a}
    \end{subfigure}
    ~ 
    \begin{subfigure}[b]{0.4\textwidth}
        \includegraphics[width=\textwidth]{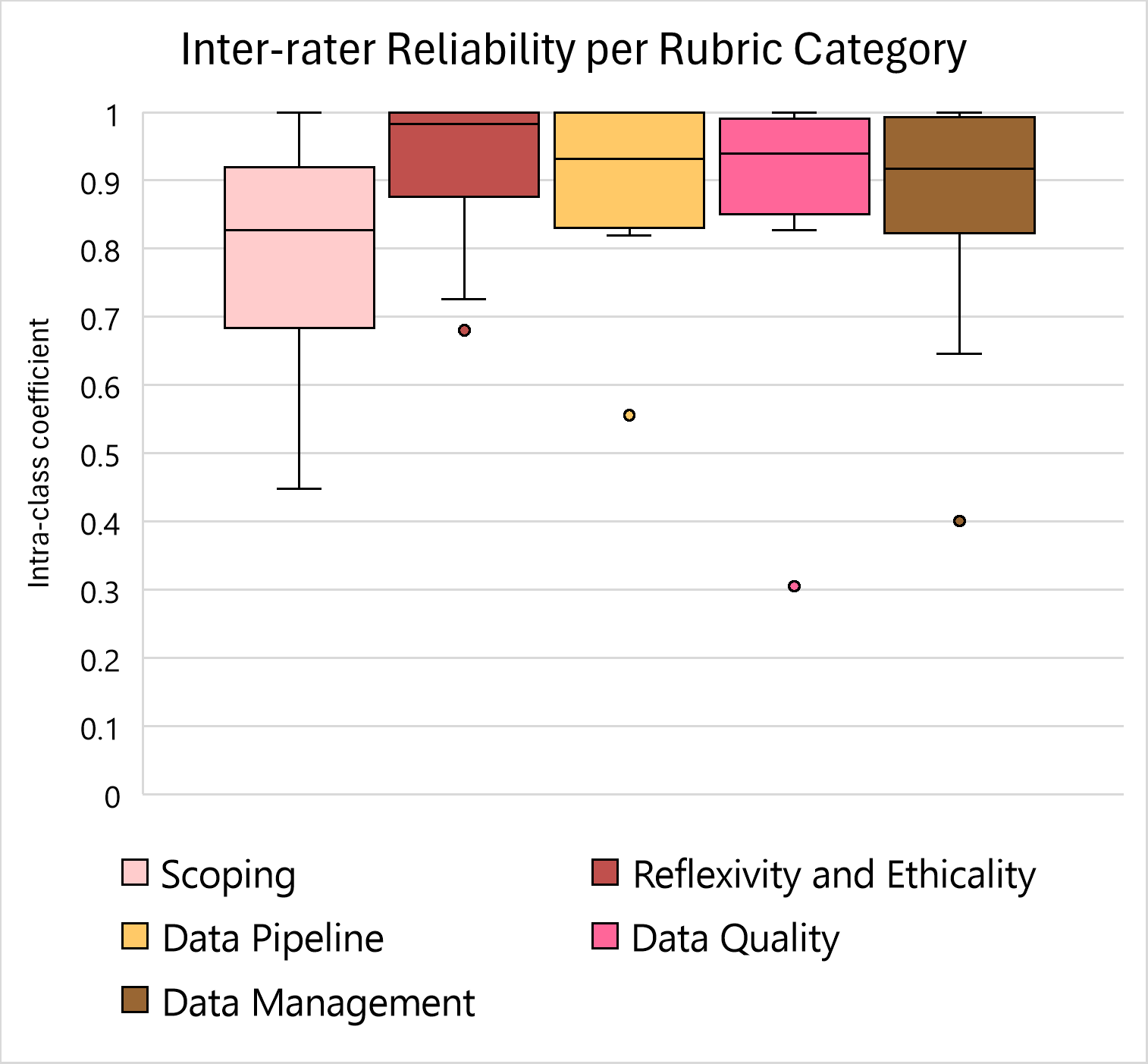}
        \caption{IRR for rubric categories in round 5}
        \label{fig:fig1b}
    \end{subfigure}
    \caption{Inter-rater reliability (IRR) \textbf{(a)} Across evaluation rounds,  and \textbf{(b)} Within round 5 across rubric categories. Improvement of IRR across rounds and ultimate high IRR across categories provides evidence that the multi-stage quality and consistency process described in Sec. \ref{methods} was successful. In addition to this quantitative measure, we conducted qualitative participatory evaluations with reviewers in each round; see \textbf{R1} and Appendix.
}\label{fig:fig1}
\end{figure*}

\textbf{R1. Inter-rater reliability suggests the evaluations are consistent and reliable.} We observed a quantifiable improvement in IRR per dataset across successive evaluation rounds. The ICC values progressively increased moving from ``fair'' to ``excellent'' agreement among raters. In the final round, the median ICC value for the 30 datasets evaluated was 0.90 (Figure 1\subref{fig:fig1a}). Despite the high level of qualitative human interpretation present when evaluating IRR across elements as compared to datasets, the final round had very high agreement, with ICC values with medians ranging from 0.83-0.98 across rubric categories (Figure 1\subref{fig:fig1b}). The improvements in IRR confirm the effectiveness of our iterative refinement approach. By continuously enhancing the rubric and its guidelines, we achieved a high level of consistency in evaluations, demonstrating the rubric's potential as a reliable tool for assessing dataset documentation in machine learning. This high level of agreement underscores the clarity and effectiveness of the rubric's criteria in guiding evaluators to consistent outcomes. These findings are critical as they establish the rubric's credibility and pave the way for its broader application and acceptance within the ML community. Additionally, the findings demonstrate the utility and rigor of qualitative human evaluations. ICC values for each of 5 rounds is shown in Figure 1\subref{fig:fig1a} and across rubric categories in Figure 1\subref{fig:fig1b}. Additional results are provided in the Appendix. 

\begin{figure*}
    \centering
    \begin{subfigure}[b]{0.49\textwidth}
        \includegraphics[width=7cm,height=5cm]{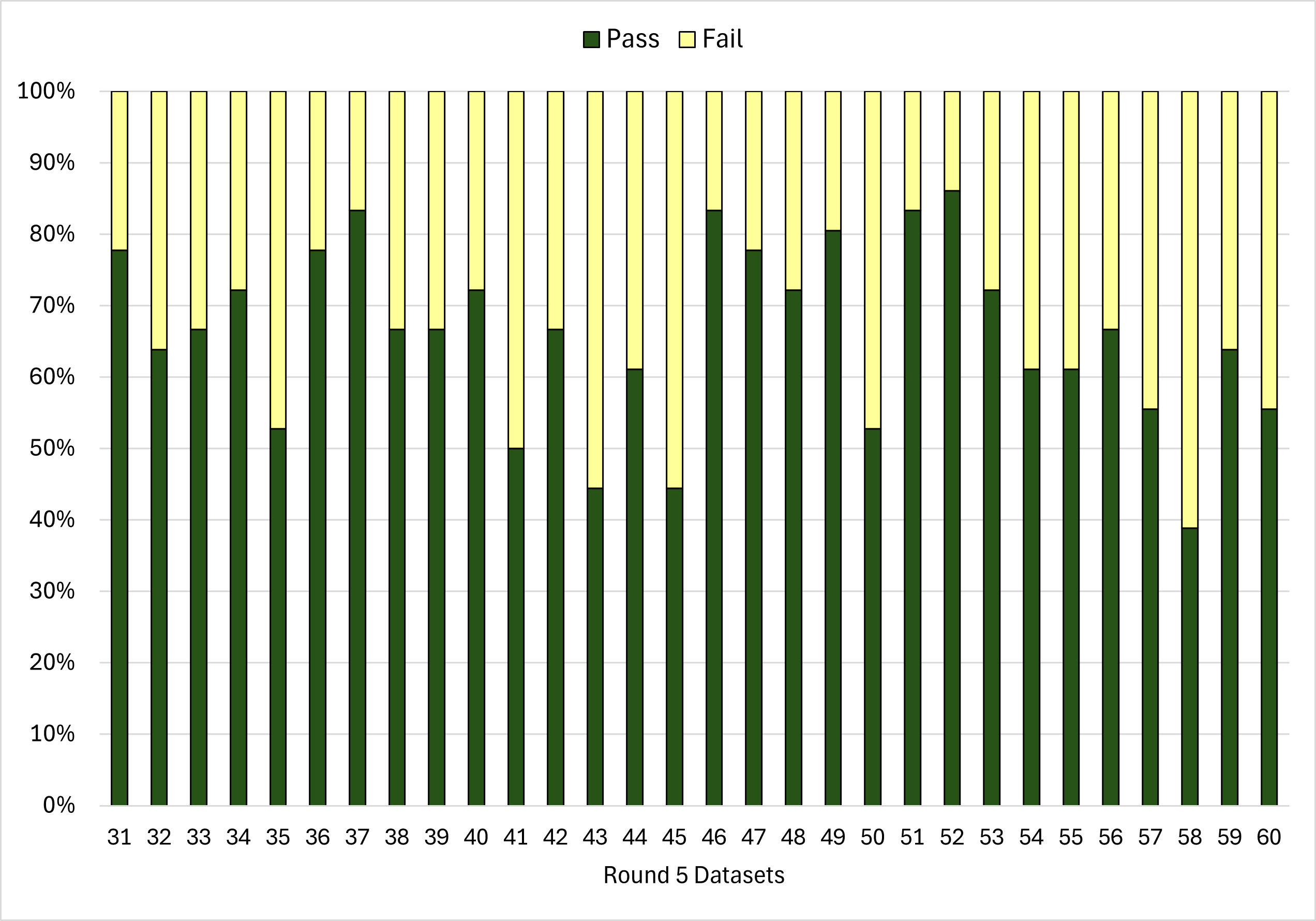}
        \caption{Per dataset, Minimum standard}
        \label{fig:fig2a}
    \end{subfigure}
    ~ 
    \begin{subfigure}[b]{0.49\textwidth}
        \includegraphics[width=7cm,height=5cm]{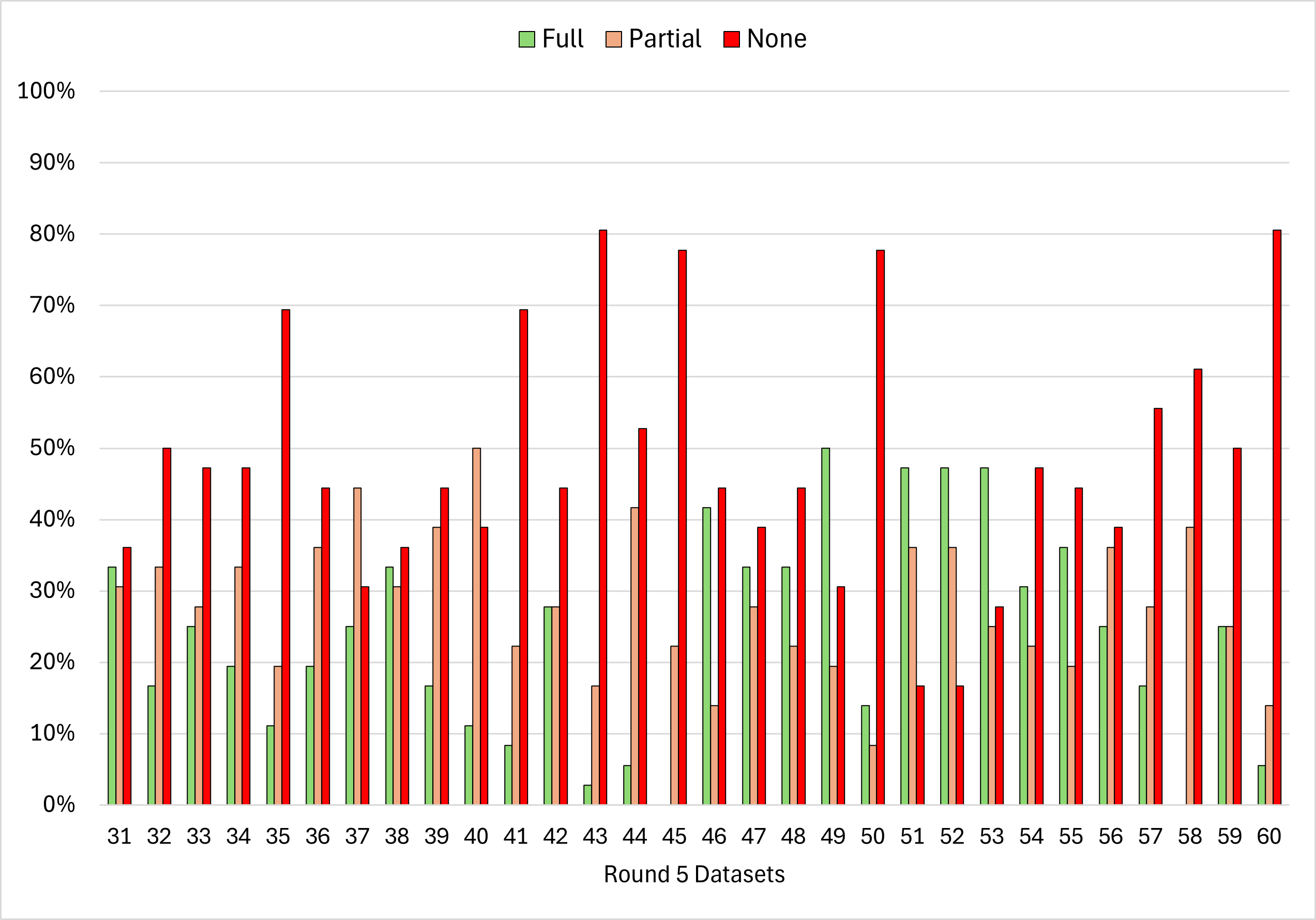}
        \caption{Per dataset, Standard of excellence}
        \label{fig:fig2b}
    \end{subfigure}
    ~
    \begin{subfigure}[b]{0.49\textwidth}
        \includegraphics[width=7cm,height=5cm]{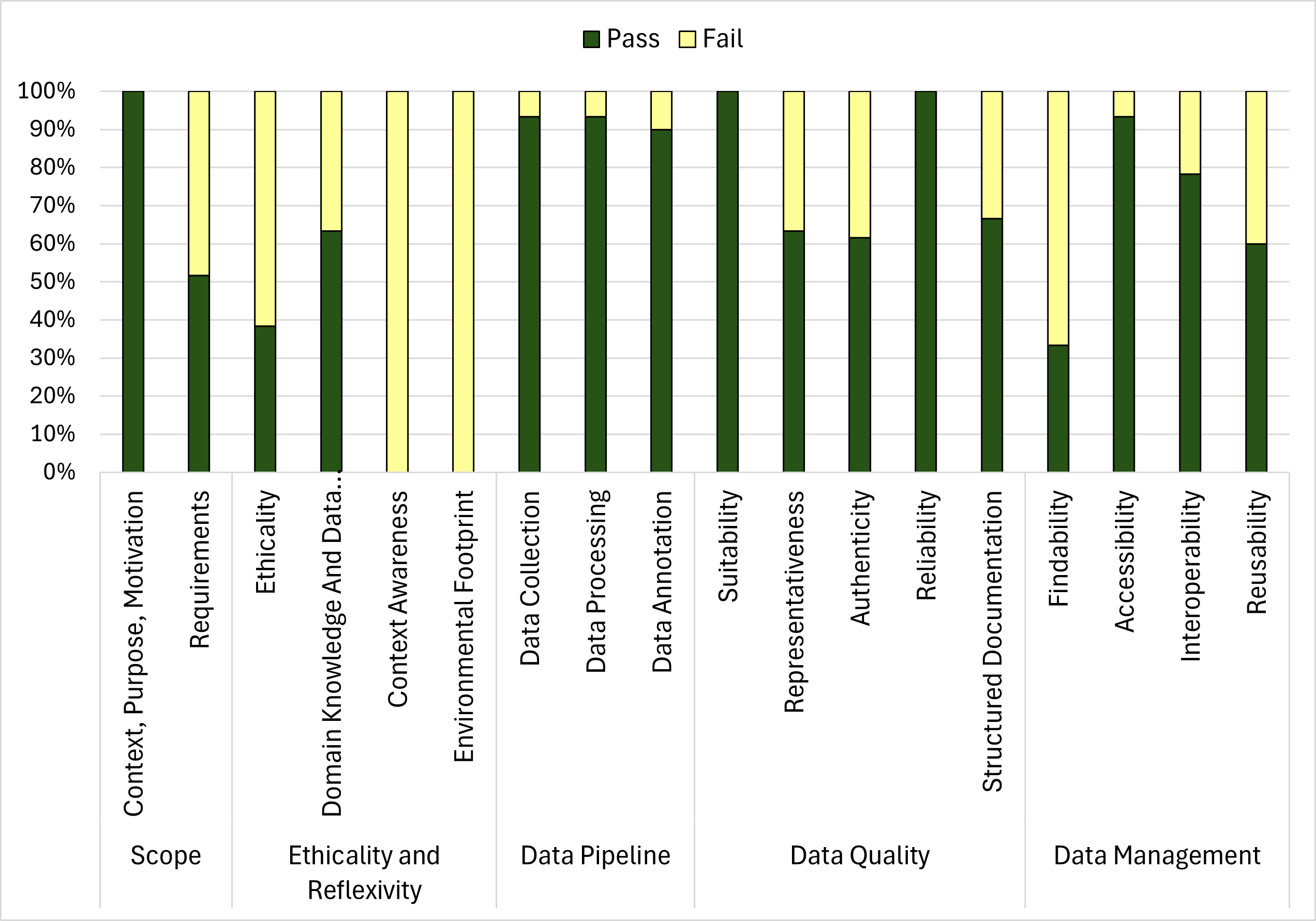}
        \caption{Per element, minimum standard}
        \label{fig:fig2c}
    \end{subfigure}
    ~
    \begin{subfigure}[b]{0.49\textwidth}
        \includegraphics[width=7cm,height=5cm]{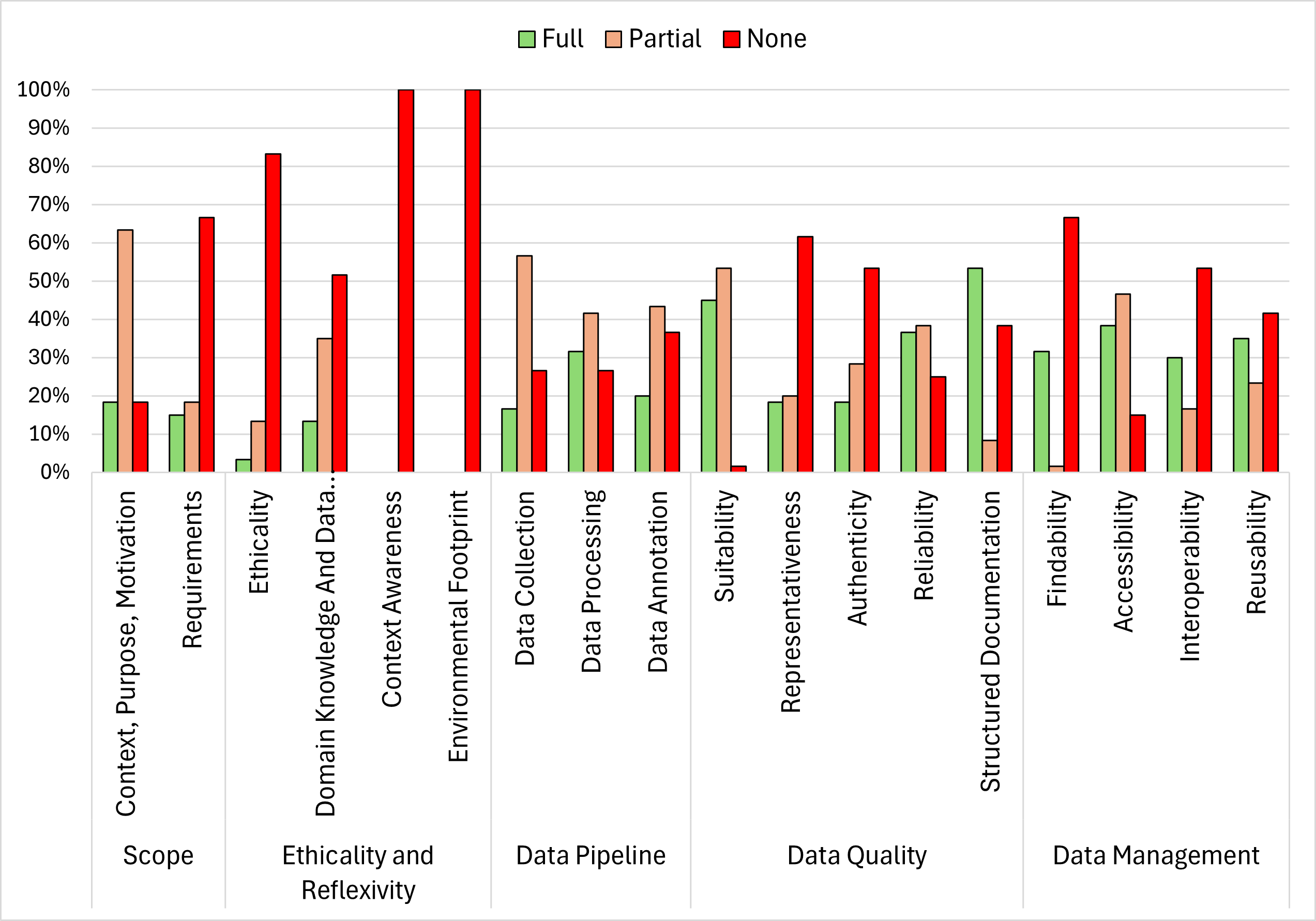}
        \caption{Per element, Standard of excellence}
        \label{fig:fig2d}
    \end{subfigure}
    \caption{Percentage of completed documentation per dataset (a,b) and per element (c,d) in round 5 (i.e. after a multi-step iterative process to improve quality). In \textbf{(a)} we observe that the highest scoring dataset fulfilled 86\% of criteria to meet the minimum standard of quality while the lowest fulfilled only 39\%; in \textbf{(b)} for the standard of excellence we see similar spread (approximately 50\% difference but lower attainment (highest fulfilled 50\% of criteria, lowest two fulfill none of the criteria for excellence); see \textbf{R2}. In both \textbf{(c)} minimum standard and \textbf{(d)} excellence we observe that those elements more closely related to model-work (such as `suitability' and `reliability') are more consistently fulfilled; see \textbf{R3}.
    }
    \label{fig:fig2}
    \vspace{-4mm}
\end{figure*}

\textbf{R2. Documentation quality varies widely across datasets.} To gauge the extent of documentation provided, we calculated for each dataset the percentage of rubric elements that received a `pass' and `fail' score for the minimum standard and a `full', `partial', and `none' score for the standard of excellence. Since each dataset was evaluated by 2 reviewers and the rubric consisted of 18 elements, we averaged the score across both reviewers and divided by 18. The results are shown in Fig 2\subref{fig:fig2a} and 2\subref{fig:fig2b}. Across all datasets evaluated during the final round, one fulfilled 86\% of the minimum standard criteria (highest pass rate), while another fulfilled only 39\% (lowest pass rate), a 47\% difference. The absolute difference between the best and worst performing papers at the standard of excellence criteria is similar, with the best-performing paper scoring `full' on 50\% of the standard of excellence and the two worst-performing receiving no `full' scores. These results demonstrate that documentation varies widely across datasets and there is great scope for improvement in documentation practices from a data curation lens, particularly to meet a standard of excellence. 

\textbf{R3. NeurIPS prioritizes model-work adjacent documentation.} To analyze where data practices could be improved, we measured the completion of documentation for each rubric element and category by calculating the number of `pass', `fail' scores and `full', `partial', `none' scores for all 60 evaluations (2 per each dataset in round 5) and divided the number by 60. Notably, NeurIPS papers tended to perform better at certain rubric elements than at others (see Figure 2\subref{fig:fig2c}, 2\subref{fig:fig2d}). Documentation for the minimum standard for `context, purpose, motivation', `suitability', and `reliability' was 100\% fulfilled. Additionally, all 3 elements in the data pipeline category were 93\% fulfilled. This highlights the areas that are prioritized and considered primary for documentation by dataset creators. These are also areas that are standard to report for publication. NeurIPS has also been able to guide and encourage greater focus through the suggested submission requirements for `structured documentation' (i.e., datasheets \cite{gebru_datasheets_2021}, data statements \cite{bender_data_2018}, etc.) and some aspects of the data management rubric elements. For example, documentation for a dataset clearly stated the problem domain of NLP and computer vision and the relevance of the new dataset being introduced in creating speech-based  rather than text-based input for assistive devices (`context, purpose, motivation'), discussed the feasibility of their dataset (`suitability'), and provided a datasheet (`structured documentation'). 

\textbf{R4. Documentation is rarely context-aware and typically does not quantify environmental footprint.} The rubric elements with the worst performance across round 5 evaluations are `context awareness' and `environmental footprint', both with 0\% pass rates of the minimum standard (and subsequently of the standard of excellence). Papers fail the `context awareness' rubric criteria by not including a dedicated positionality statement (a statement of authors' institutional affiliations is not considered as a statement of or reflection on positionality). For the standard of excellence, less than 20\% of papers receive a `full' or `partial' score for the `ethicality' standard of excellence. That is: even those papers that make use of the proportionality principle and document informed consent tend to do so only as much as required by ethics checklists, with additional ethical discussion rarely included. The evaluated datasets also fail the `environmental footprint' criteria because none of them quantitatively assess the environmental footprint associated with dataset creation.

\textbf{R5. Documentation often remains incomplete.} The results indicate that even for those datasets with well-documented elements for the minimum standard, rigorous documentation is ultimately lacking. For example, in the case of `reliability', papers tend to pass the minimum standard by describing the phenomena represented by the data (e.g., describing the videos from which screen capture data were generated), but fail the standard of excellence by not providing a mechanism by which others could verify what was being represented (e.g., no way for anyone else to access the videos used to produce screen captures). As in the case of `reliability', and as intended in the rubric's design, papers perform better at the minimum standard than at the standard of excellence: 14 out of 18 rubric criteria have a minimum standard pass rate over 50\%, compared to 1 of 18 with `full' scores over 50\% and 3 of 18 with `partial' scores over 50\% for the standard of excellence. 



\begin{figure*}[h!]
    \centering
    \begin{subfigure}[b]{0.4\textwidth}
        \includegraphics[width=\textwidth]{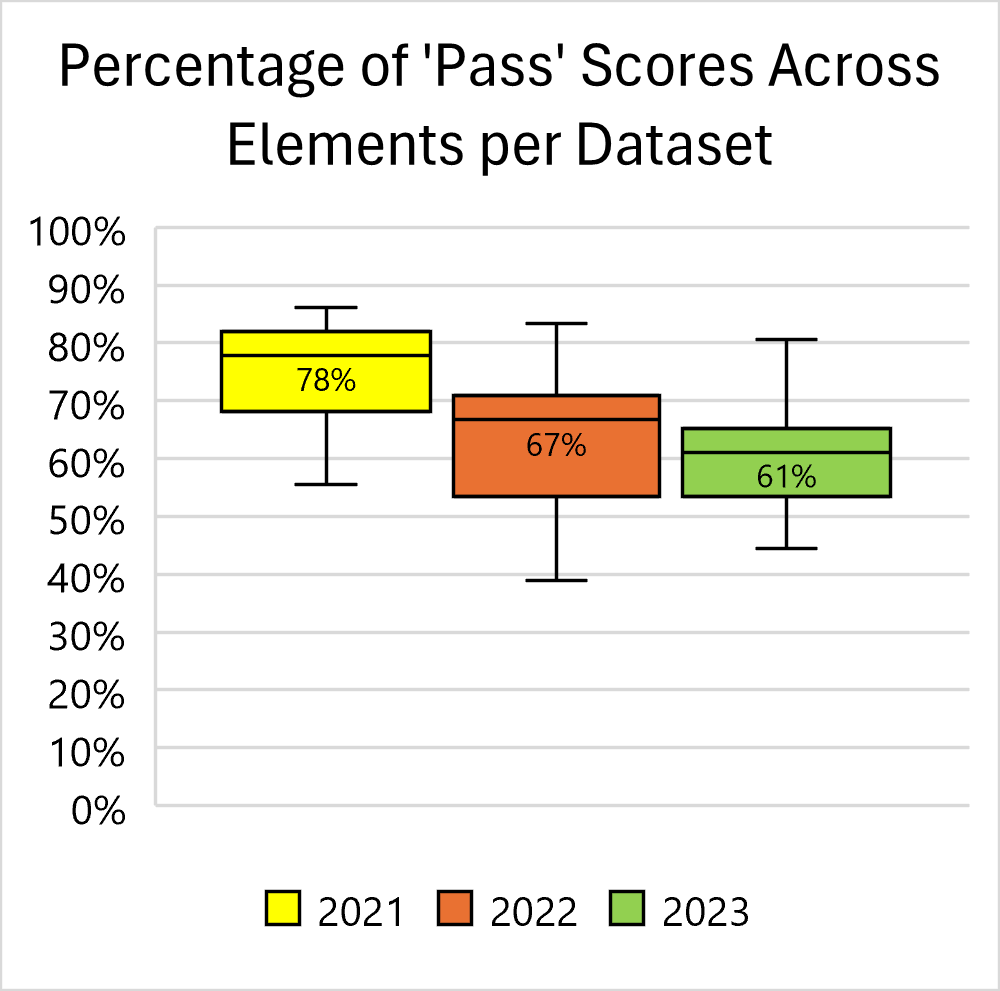}
        \caption{Minimum standard}
        \label{fig:fig3a}
    \end{subfigure}
    ~ 
    \begin{subfigure}[b]{0.4\textwidth}
        \includegraphics[width=\textwidth]{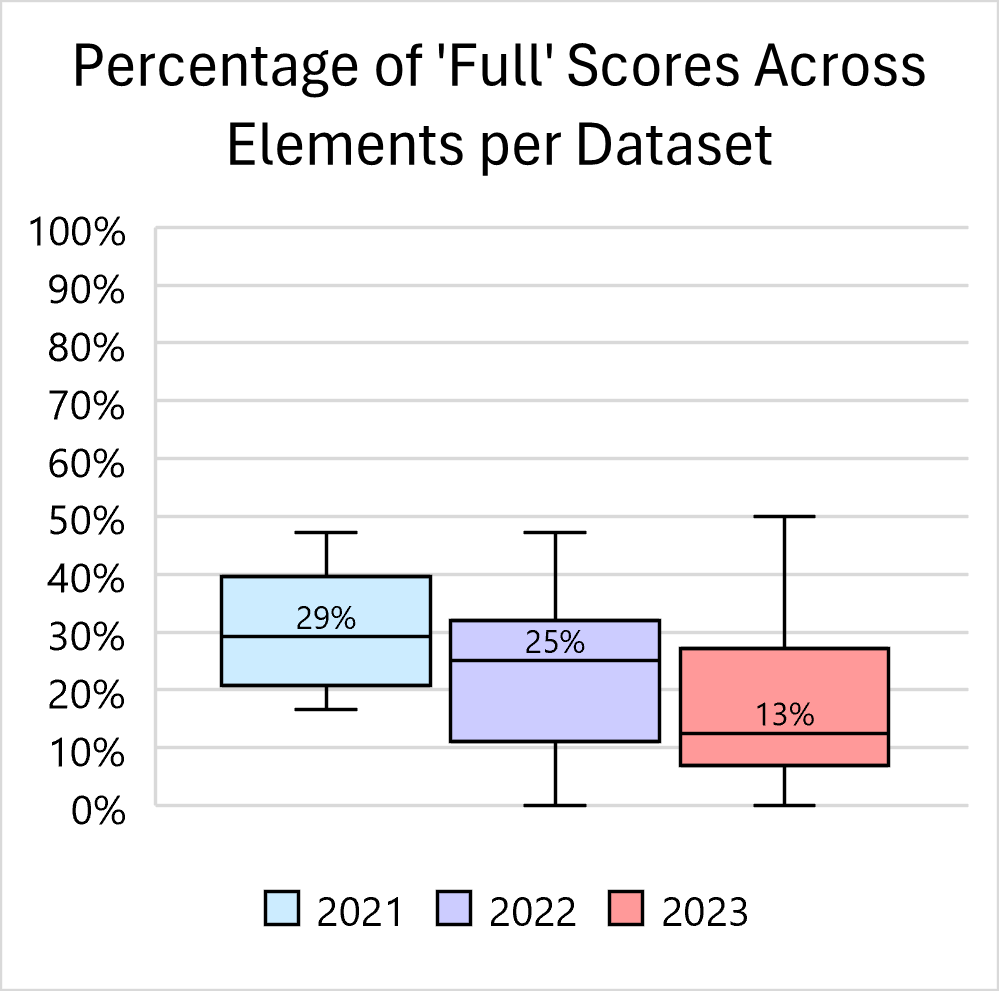}
        \caption{Standard of excellence}
        \label{fig:fig3b}
    \end{subfigure}
    \caption{Temporal distribution across years 2021-2023, \textbf{(a)} `pass scores' for the minimum standard of quality and \textbf{(b)} `full scores' for the standard of excellence across elements. In both cases there is no change across time; see \textbf{R6}.
}\label{fig:fig3}
\vspace{-4mm}
\end{figure*}

\textbf{R6. Findings suggest no improvements occurred over time.} We evaluated an even distribution of datasets published in 2021, 2022, and 2023 for round 5. Figure \ref{fig:fig3} shows the results of the percentage of `pass' and `full' scores across elements for each dataset summarized by year. Particularly, we can observe a slight downward trend in documentation scores across the years evaluated: from 2021 to 2023, the median percentage of `pass' scores per dataset for the minimum standard goes from 78\%, to 67\%, to 61\%, while the median percentage of `full' scores per dataset for the standard of excellence goes from 29\%, 25\%, and 13\%, respectively. In 2021, the call for papers for the D\&B track required the submission of dataset documentation, URL for accessing the dataset, details about data licensing, hosting, and maintenance, and for authors to ensure easy reproducibility \cite{noauthor_call_2021}. In the following years, additional requirements around datasets being in widely used formats, long-term preservation, inclusion of and access to metadata, and usage of persistent identifiers were added \cite{noauthor_call_2022,noauthor_call_2023}. Despite the increasing stringency of requirements, we could not find any evidence in this sample to suggest that the extent of provided documentation improved over time, pointing to the need for an encompassing structure and framework by which to assess documentation practices and pinpoint areas of improvement. Furthermore, the use of structured documentation also reduced over time, from 80\% in 2021 to 50\% in 2023. 


\section{Discussion} \label{discusion}

\subsection{How to Improve Data Curation at NeurIPS: Strategies and Resources} \label{discuss_resources}

Our findings identify areas for which datasets have low or no documentation. To aid dataset creators in strengthening their curation processes, we recommend the adoption of the following methods and resources, particularly for the elements `requirements', `ethicality', `context awareness', `environmental footprint', `findability', and `reusability'. 

First, to address \textbf{`requirements'} we echo recommendations regarding the creation of \textbf{purpose statements} \cite{andrews_ethical_2023}. Stating how dataset creators translated the ``real-world'' problem they are addressing into a ``ML problem'' for which the dataset is created \cite{muller_forgetting_2022,passi_problem_2019} promotes transparency. This process consists of numerous decisions, expertise, and assumptions that should be documented in order to understand the context in which the problem situation was framed. In cases of harmful ML models, it has been seen that this translation process can lead to the creation of bad proxy data that unfairly represents the real-world scenario \cite{oneil_weapons_2017}. Furthermore, sharing this process reveals the practicalities of performing data work which is often hidden and considered under-valued \cite{heger_understanding_2022,thomer_craft_2022}. Particularly, it is important for dataset creators to distinguish between the \textbf{initial formulation} of the problem \textbf{vs.} the \textbf{dataset creation scheme} detailing how the dataset development was actually executed. The latter is often documented after the development of the data and can be impaired by the ``forgetting practice'' of only recording conclusions \cite{muller_forgetting_2022}. As Muller et al. distinctly point out, ``Measurement plans tend to record conclusions, not rationales… Other people then work with those conclusions, and have no way to access those unrecorded rationales.'' \citep[p.~9]{muller_forgetting_2022}.

We urge dataset creators to explicitly document how the benefits of developing their dataset outweigh the harms of creating it to improve reflection on \textbf{`ethicality'}. In other words, the \textbf{proportionality principle} must be considered. In ethics, it is understood that actions have positive and negative effects simultaneously. This is called the double effect. ``Applications of double effect always presuppose that some kind of proportionality condition has been satisfied. Traditional formulations of the proportionality condition require that the value of promoting the good end outweigh the disvalue of the harmful side effect.'' \cite{mcintyre_doctrine_2023}. The submission checklist for NeurIPS requires authors to document potential positive and negative societal impacts of their work; we support this and additionally recommend the checklist be amended to encourage comparison and reflection on these in proportion to each other, which very few datasets from our evaluations explicitly do.

\textbf{`Context awareness' }was not well demonstrated in any dataset we evaluated, correlated with the lack of positionality statements in the NeurIPS D\&B track as a whole. Past research from the track has pointed to the importance of annotator \textbf{positionality}, as researchers' social identity and implicit biases impact data-related choices \cite{andrews_ethical_2023}; this recommendation is also made from a feminist HCI (human-computer interaction) perspective with the goal to increase \textbf{reflexivity} in ML dataset development \cite{Scheuerman_Hanna_Denton_2021}. 
See Appendix for examples of positionality and reflection statements of this work.

Data curation in ML encompasses many phases, each responsible for significant emissions due to energy consumption \cite{dayarathna_data_2015,innovation_how_2020,lacoste_quantifying_2019,luccioni_counting_2023,office_of_energy_efficiency__renewable_energy_data_2023,patterson_carbon_2021}. Several datasets in our sample ranged from over one million instances \cite{islam_caesar_2022,kaltenborn_climateset_2023,saito_open_2020,xuanwen_dgraph_2022} to several billion instances \cite{dell_american_2023,gadre_datacomp_2024,hambro_dungeons_2022,penedo_refinedweb_2023}. Owing to their volume, these datasets are anticipated to have a significant environmental footprint, beyond the impacts associated with model training which have tended to be the focus in ML. There were no quantitative assessments of \textbf{`environmental footprint'} in the evaluated datasets. This quantification is crucial for several reasons: it allows for the assessment and comparison of carbon footprint across different projects, facilitating more informed decisions about resource allocation and model design \cite{luccioni_counting_2023,strubell_energy_2019}. Understanding these impacts can also drive the development of more energy-efficient algorithms and hardware, contributing to broader efforts to mitigate climate change \cite{kaack_aligning_2022,lacoste_quantifying_2019,thompson_computational_2020}. Moreover, as public awareness of environmental issues grows, the ML community faces increasing pressure to demonstrate accountability and progress toward sustainability \cite{kaack_aligning_2022,patterson_carbon_2021,strubell_energy_2019}. 
Transparent reporting of carbon emissions can enhance the credibility of research institutions and companies in the field. By understanding where these emissions originate, researchers and engineers can better target interventions, such as optimizing algorithms for efficiency or sourcing providers with renewable energy for power-intensive tasks, to mitigate the ecological consequences of their work \cite{kaack_aligning_2022, lacoste_quantifying_2019, thompson_computational_2020}. 
We provide a list of strategies in the Appendix on how to report environmental footprint for dataset development. 

To improve `\textbf{findability}', we urge NeurIPS to require datasets to have \textbf{metadata} (not just data) assigned a persistent identifier and hosted in a searchable repository (such as Zenodo). While we recognize that having this requirement for all datasets may be infeasible due to volume, sensitivity, and other factors, having this information \textit{for metadata} will help provide information about the dataset that will be available in the long-term, even if the data are gone. Although many datasets are provided on GitHub or platforms hosted by the dataset creators, these URLs suffer from a high likelihood of \textbf{link rot }(that the link will no longer be available over time) \cite{klein_scholarly_2014}. Lastly, to improve `\textbf{reusability}', we echo the inclusion of identifier information, dataset characteristics, and dataset provenance as outlined in The Data Provenance Initiative \cite{longpre_data_2023}. 

\subsection{What Next: A Proposal for Peer-Review} \label{discuss_proposal}

Peer-review processes are constantly progressing with evolving needs,  and they require multiple lenses \cite{NeurIPS_changestoreview_2021,NeurIPS_retrospectiveethics_2021}. We propose that our evaluation framework can provide a structure that enhances the submission and peer-review process for the NeurIPS Datasets and Benchmarks track. Dataset creators, by applying the rubric and associated resources, would be more easily able to conduct their dataset development processes with data curation in mind. On the other side, dataset reviewers can use the rubric to evaluate dataset documentation, highlight targeted areas of improvement where data curation standards can be applied, and provide  recommendations. The framework speaks directly to the evaluation of documentation required from peer-reviewers in the review form - ``Documentation: For datasets, is there sufficient detail on data collection and organization, availability and maintenance, and ethical and responsible use? Note that dataset submissions should include documentation and intended uses; a URL for reviewer access to the dataset; and a hosting, licensing and maintenance plan.'' \cite{NeurIPSDBtrack_reviewerguidelines}. 
The presented framework allows for the systematic, precise, and encompassing evaluation of these details beyond the prompts present in the current review form, through the criteria presented for the elements in the data pipeline category (i.e., data collection and organization), the data management category based on the FAIR principles \cite{wilkinson_fair_2016} (i.e., availability and maintenance), and the ethicality and reflexivity category (i.e., ethical and responsible use). 
The use of this framework would also enable reviewers to consistently determine whether an additional ethics review is needed. 
A past consistency experiment for peer-review at NeurIPS in 2014
showed that ``if the conference reviewing had been run with a different committee, only half of the papers presented at the conference would have been the same'' \citep[p.~3]{Cortes_Lawrence_2021}. The 2021 version of the experiment was ``...consistent with the 2014 experiment when the conference was an order of magnitude smaller.'' \cite{Alina_2021}. Our framework can thus help provide a means of consistency in terms of dataset documentation. We suggest incorporating the framework by introducing a dedicated `dataset documentation' reviewer role to the NeurIPS D\&B track. This can initially be similar to the ethics reviewer, who performs reviews only for those papers that are flagged, but may later evolve to be a part of the core peer-review team. We recommend a dedicated reviewer because of the relatively intense process of evaluating each dataset and the specialized skillset that it will require. Although we found that the time requirement to evaluate each dataset ranged from 35 minutes to 2 hours, the average time  in later rounds was limited to 1 hour. 
Additionally, as with ethics reviews, the results from the dataset documentation review should advise the acceptance of the paper as poor documentation ultimately leads to poor reusability and reproducibility which would defeat the purpose of the D\&B track. 

\subsection{Limitations} \label{discuss_limits}

We identify five limitations of our work. \textbf{1.} The results we showcase are based on the sample set of datasets we evaluated. Although these datasets were randomly chosen and evenly distributed between 2021-2023, there is a chance for the selected datasets to be unrepresentative of the datasets and benchmarks published at the NeurIPS D\&B track as a whole. 
\textbf{2.} Our analysis is based on descriptive and interpretative evaluations completed by a mix of reviewers. Although we took careful steps in our iterative process to clarify and normalize standards of interpretation, all results are contingent on the human processes of differing evaluation styles. As with all peer-review, the results are thus dependent on the reviewers. \textbf{3.} It is understood that a given reviewer would evaluate each element in the rubric similarly across all datasets. However, we also conducted a disagreement review. This means that a reviewer could change their perspective for a specific element based on the comments of another reviewer, if a disagreement was flagged. This does \textit{not} mean that the reviewer would then update their scores and comments for that element for all \textit{other} datasets they evaluated. Thus some inconsistencies in how the one element is evaluated across all the datasets might be introduced as the cost of improving within-dataset review quality. We believe the impact of this limitation is low, as our results showed minimal disagreement in the final round (see Figure 1\subref{fig:fig1a}).
\textbf{4.} Our rubric is designed to enable qualitative and holistic evaluation of each paper on a standardized basis. We believe strongly in the merits of this approach, however it does limit the amount of insight we can have about how properties of the dataset itself influence curation practices. An interesting complementary future approach to study this would be to code characteristics of datasets and see if the trends we identify vary when decomposed by these characteristics, i.e., whether there are specific documentation trends across types of ML datasets or metadata.
\textbf{5.} Using documentation to understand the curation process is not a substitute for being directly involved in the process or communicating with the dataset creators. As such, it is possible and ultimately a limitation that the reality of data curation is more complex than what is covered in documentation or our framework. In such cases, things can become overly simplified in the documentation and auditing process, e.g., box ticking instead of genuine reflection and evaluation. An ethnographic study of data curation would yield different, additional, insights that we cannot provide. This is currently a limitation and opportunity for future work.

\section{Conclusion} \label{conclusion}

By giving datasets and benchmarks a dedicated venue, NeurIPS has sent a clear message that dataset quality is the foundation of continued progress in ML applications and fundamental understanding. There is no better database of knowledge than data curation to aid in this venture. Our evaluation framework adopts concepts from these fields for ML and provides a practical lens on how NeurIPS can spearhead the requirement for rigorous data curation in ML. The enhancements due to the framework are designed to improve the quality, ethicality, and human oversight of new datasets and benchmarks, fostering greater scientific integrity and advancement. 

\begin{ack}
This research was partially supported by NSERC through RGPIN-2016-06640 and by the Canada Foundation for Innovation and the Ontario Research Fund.
\end{ack}
\newpage

\bibliographystyle{ACM-Reference-Format}
\bibliography{DCML_neurips24}

\newpage
\section*{NeurIPS Paper Checklist}

\begin{enumerate}

\item {\bf Claims}
    \item[] Question: Do the main claims made in the abstract and introduction accurately reflect the paper's contributions and scope?
    \item[] Answer: \answerYes{} 
    \item[] Justification: We claim that assessing dataset development from a data curation lens can improve the documentation practices in the NeurIPS Datasets and Benchmarks track. We contribute an evaluation framework to support this claim. The abstract and introduction provide further details on our paper’s contributions and scope.

\item {\bf Limitations}
    \item[] Question: Does the paper discuss the limitations of the work performed by the authors?
    \item[] Answer: \answerYes{} 
    \item[] Justification: We outline the limitations of our methods and resulting findings in Section \ref{discuss_limits}.

\item {\bf Theory Assumptions and Proofs}
    \item[] Question: For each theoretical result, does the paper provide the full set of assumptions and a complete (and correct) proof?
    \item[] Answer: \answerNA{} 
    \item[] Justification: The paper does not include theoretical results.

    \item {\bf Experimental Result Reproducibility}
    \item[] Question: Does the paper fully disclose all the information needed to reproduce the main experimental results of the paper to the extent that it affects the main claims and/or conclusions of the paper (regardless of whether the code and data are provided or not)?
    \item[] Answer: \answerYes{} 
    \item[] Justification: We do not perform experiments in the traditional sense, but we provide methods to reproduce the results provided in the paper. 

\item {\bf Open access to data and code}
    \item[] Question: Does the paper provide open access to the data and code, with sufficient instructions to faithfully reproduce the main experimental results, as described in supplemental material?
    \item[] Answer: \answerNA{} 
    \item[] Justification: The paper does not include experiments. However the rubric evaluations and analyses are hosted on Zenodo.

\item {\bf Experimental Setting/Details}
    \item[] Question: Does the paper specify all the training and test details (e.g., data splits, hyperparameters, how they were chosen, type of optimizer, etc.) necessary to understand the results?
    \item[] Answer: \answerNA{} 
    \item[] Justification: The paper does not include experiments. 

\item {\bf Experiment Statistical Significance}
    \item[] Question: Does the paper report error bars suitably and correctly defined or other appropriate information about the statistical significance of the experiments?
    \item[] Answer: \answerNA{} 
    \item[] Justification: The paper does not include experiments. 

\item {\bf Experiments Compute Resources}
    \item[] Question: For each experiment, does the paper provide sufficient information on the computer resources (type of compute workers, memory, time of execution) needed to reproduce the experiments?
    \item[] Answer: \answerNA{} 
    \item[] Justification: The paper does not include experiments.
    
\item {\bf Code Of Ethics}
    \item[] Question: Does the research conducted in the paper conform, in every respect, with the NeurIPS Code of Ethics \url{https://neurips.cc/public/EthicsGuidelines}?
    \item[] Answer: \answerYes{} 
    \item[] Justification: The research conforms with the NeurIPS Code of Ethics. 

\item {\bf Broader Impacts}
    \item[] Question: Does the paper discuss both potential positive societal impacts and negative societal impacts of the work performed?
    \item[] Answer: \answerNA{} 
    \item[] Justification: Our work does not present a dataset or benchmark. Our work provides a framework to increase accountability, transparency, reusability, and reproducibility of ML datasets. 
    
\item {\bf Safeguards}
    \item[] Question: Does the paper describe safeguards that have been put in place for responsible release of data or models that have a high risk for misuse (e.g., pretrained language models, image generators, or scraped datasets)?
    \item[] Answer: \answerNA{} 
    \item[] Justification: Our paper does not pose any such risks.

\item {\bf Licenses for existing assets}
    \item[] Question: Are the creators or original owners of assets (e.g., code, data, models), used in the paper, properly credited and are the license and terms of use explicitly mentioned and properly respected?
    \item[] Answer: \answerNA{} 
    \item[] Justification: The paper does not use existing assets.

\item {\bf New Assets}
    \item[] Question: Are new assets introduced in the paper well documented and is the documentation provided alongside the assets?
    \item[] Answer: \answerYes{} 
    \item[] Justification: We introduce new assets in the form of evaluations of datasets using our framework. The process of performing these evaluations is discussed in Section \ref{methods}.

\item {\bf Crowdsourcing and Research with Human Subjects}
    \item[] Question: For crowdsourcing experiments and research with human subjects, does the paper include the full text of instructions given to participants and screenshots, if applicable, as well as details about compensation (if any)? 
    \item[] Answer: \answerNA{} 
    \item[] Justification: The paper does not involve crowdsourcing nor research with human subjects.

\item {\bf Institutional Review Board (IRB) Approvals or Equivalent for Research with Human Subjects}
    \item[] Question: Does the paper describe potential risks incurred by study participants, whether such risks were disclosed to the subjects, and whether Institutional Review Board (IRB) approvals (or an equivalent approval/review based on the requirements of your country or institution) were obtained?
    \item[] Answer: \answerNA{} 
    \item[] Justification: The paper does not involve crowdsourcing nor research with human subjects.

\end{enumerate}



\section*{Supplementary material}

Answer: [N/A]

Justification: The paper does not include theoretical results.

Guidelines:

- The answer NA means that the paper does not include theoretical results.
- All the theorems, formulas, and proofs in the paper should be numbered and cross-referenced.
- All assumptions should be clearly stated or referenced in the statement of any theorems.
- The proofs can either appear in the main paper or the supplemental material, but if they appear in the supplemental material, the authors are encouraged to provide a short proof sketch to provide intuition.
- Inversely, any informal proof provided in the core of the paper should be complemented by formal proofs provided in appendix or supplemental material.
- Theorems and Lemmas that the proof relies upon should be properly referenced.

4. **Experimental Result Reproducibility**

Question: Does the paper fully disclose all the information needed to reproduce the main experimental results of the paper to the extent that it affects the main claims and/or conclusions of the paper (regardless of whether the code and data are provided or not)?

Answer: [Yes]

Justification: We do not perform experiments in the traditional sense, but we provide methods to reproduce the results provided in the paper.

Guidelines:

- The answer NA means that the paper does not include experiments.
- If the paper includes experiments, a No answer to this question will not be perceived well by the reviewers: Making the paper reproducible is important, regardless of whether the code and data are provided or not.
- If the contribution is a dataset and/or model, the authors should describe the steps taken to make their results reproducible or verifiable.
- Depending on the contribution, reproducibility can be accomplished in various ways. For example, if the contribution is a novel architecture, describing the architecture fully might suffice, or if the contribution is a specific model and empirical evaluation, it may be necessary to either make it possible for others to replicate the model with the same dataset, or provide access to the model. In general. releasing code and data is often one good way to accomplish this, but reproducibility can also be provided via detailed instructions for how to replicate the results, access to a hosted model (e.g., in the case of a large language model), releasing of a model checkpoint, or other means that are appropriate to the research performed.
- While NeurIPS does not require releasing code, the conference does require all submissions to provide some reasonable avenue for reproducibility, which may depend on the nature of the contribution. For example
  (a) If the contribution is primarily a new algorithm, the paper should make it clear how to reproduce that algorithm.
  (b) If the contribution is primarily a new model architecture, the paper should describe the architecture clearly and fully.
  (c) If the contribution is a new model (e.g., a large language model), then there should either be a way to access this model for reproducing the results or a way to reproduce the model (e.g., with an open-source dataset or instructions for how to construct the dataset).
  (d) We recognize that reproducibility may be tricky in some cases, in which case authors are welcome to describe the particular way they provide for reproducibility. In the case of closed-source models, it may be that access to the model is limited in some way (e.g., to registered users), but it should be possible for other researchers to have some path to reproducing or verifying the results.

5. **Open access to data and code**

Question: Does the paper provide open access to the data and code, with sufficient instructions to faithfully reproduce the main experimental results, as described in supplemental material?

Answer: [N/A]

Justification: The paper does not include experiments. However the rubric evaluations and analyses are hosted on Zenodo.

Guidelines:

- The answer NA means that paper does not include experiments requiring code.
- Please see the NeurIPS code and data submission guidelines (`https://nips.cc/public/guides/CodeSubmissionPolicy`) for more details.
- While we encourage the release of code and data, we understand that this might not be possible, so "No" is an acceptable answer. Papers cannot be rejected simply for not including code, unless this is central to the contribution (e.g., for a new open-source benchmark).
- The instructions should contain the exact command and environment needed to run to reproduce the results. See the NeurIPS code and data submission guidelines (`https://nips.cc/public/guides/CodeSubmissionPolicy`) for more details.
- The authors should provide instructions on data access and preparation, including how to access the raw data, preprocessed data, intermediate data, and generated data, etc.
- The authors should provide scripts to reproduce all experimental results for the new proposed method and baselines. If only a subset of experiments are reproducible, they should state which ones are omitted from the script and why.
- At submission time, to preserve anonymity, the authors should release anonymized versions (if applicable).
- Providing as much information as possible in supplemental material (appended to the paper) is recommended, but including URLs to data and code is permitted.

6. **Experimental Setting/Details**

Question: Does the paper specify all the training and test details (e.g., data splits, hyper-parameters, how they were chosen, type of optimizer, etc.) necessary to understand the results?

Answer: [N/A]

Justification: The paper does not include experiments.

Guidelines:

- The answer NA means that the paper does not include experiments.
- The experimental setting should be presented in the core of the paper to a level of detail that is necessary to appreciate the results and make sense of them.
- The full details can be provided either with the code, in appendix, or as supplemental material.

7. **Experiment Statistical Significance**

Question: Does the paper report error bars suitably and correctly defined or other appropriate information about the statistical significance of the experiments?

Answer: [N/A]

Justification: The paper does not include experiments.

Guidelines:

- The answer NA means that the paper does not include experiments.
- The authors should answer "Yes" if the results are accompanied by error bars, confidence intervals, or statistical significance tests, at least for the experiments that support the main claims of the paper.
- The factors of variability that the error bars are capturing should be clearly stated (for example, train/test split, initialization, random drawing of some parameter, or overall run with given experimental conditions).

- The method for calculating the error bars should be explained (closed form formula, call to a library function, bootstrap, etc.)
- The assumptions made should be given (e.g., Normally distributed errors).
- It should be clear whether the error bar is the standard deviation or the standard error of the mean.
- It is OK to report 1-sigma error bars, but one should state it. The authors should preferably report a 2-sigma error bar than state that they have a 96% CI, if the hypothesis of Normality of errors is not verified.
- For asymmetric distributions, the authors should be careful not to show in tables or figures symmetric error bars that would yield results that are out of range (e.g. negative error rates).
- If error bars are reported in tables or plots, The authors should explain in the text how they were calculated and reference the corresponding figures or tables in the text.

8. **Experiments Compute Resources**

   Question: For each experiment, does the paper provide sufficient information on the computer resources (type of compute workers, memory, time of execution) needed to reproduce the experiments?

   Answer: [N/A]

   Justification: The paper does not include experiments.

   Guidelines:

   - The answer NA means that the paper does not include experiments.
   - The paper should indicate the type of compute workers CPU or GPU, internal cluster, or cloud provider, including relevant memory and storage.
   - The paper should provide the amount of compute required for each of the individual experimental runs as well as estimate the total compute.
   - The paper should disclose whether the full research project required more compute than the experiments reported in the paper (e.g., preliminary or failed experiments that didn't make it into the paper).

9. **Code Of Ethics**

   Question: Does the research conducted in the paper conform, in every respect, with the NeurIPS Code of Ethics https://neurips.cc/public/EthicsGuidelines?

   Answer: [Yes]

   Justification: The research conforms with the NeurIPS Code of Ethics.

   Guidelines:

   - The answer NA means that the authors have not reviewed the NeurIPS Code of Ethics.
   - If the authors answer No, they should explain the special circumstances that require a deviation from the Code of Ethics.
   - The authors should make sure to preserve anonymity (e.g., if there is a special consideration due to laws or regulations in their jurisdiction).

10. **Broader Impacts**

    Question: Does the paper discuss both potential positive societal impacts and negative societal impacts of the work performed?

    Answer: [N/A]

    Justification: Our work does not present a dataset or benchmark. Our work provides a framework to increase accountability, transparency, reusability, and reproducibility of ML datasets.

    Guidelines:

    - The answer NA means that there is no societal impact of the work performed.
    - If the authors answer NA or No, they should explain why their work has no societal impact or why the paper does not address societal impact.

- Examples of negative societal impacts include potential malicious or unintended uses (e.g., disinformation, generating fake profiles, surveillance), fairness considerations (e.g., deployment of technologies that could make decisions that unfairly impact specific groups), privacy considerations, and security considerations.
- The conference expects that many papers will be foundational research and not tied to particular applications, let alone deployments. However, if there is a direct path to any negative applications, the authors should point it out. For example, it is legitimate to point out that an improvement in the quality of generative models could be used to generate deepfakes for disinformation. On the other hand, it is not needed to point out that a generic algorithm for optimizing neural networks could enable people to train models that generate Deepfakes faster.
- The authors should consider possible harms that could arise when the technology is being used as intended and functioning correctly, harms that could arise when the technology is being used as intended but gives incorrect results, and harms following from (intentional or unintentional) misuse of the technology.
- If there are negative societal impacts, the authors could also discuss possible mitigation strategies (e.g., gated release of models, providing defenses in addition to attacks, mechanisms for monitoring misuse, mechanisms to monitor how a system learns from feedback over time, improving the efficiency and accessibility of ML).

11. **Safeguards**

Question: Does the paper describe safeguards that have been put in place for responsible release of data or models that have a high risk for misuse (e.g., pretrained language models, image generators, or scraped datasets)?

Answer: [N/A]

Justification: Our paper does not pose any such risks.

Guidelines:

- The answer NA means that the paper poses no such risks.
- Released models that have a high risk for misuse or dual-use should be released with necessary safeguards to allow for controlled use of the model, for example by requiring that users adhere to usage guidelines or restrictions to access the model or implementing safety filters.
- Datasets that have been scraped from the Internet could pose safety risks. The authors should describe how they avoided releasing unsafe images.
- We recognize that providing effective safeguards is challenging, and many papers do not require this, but we encourage authors to take this into account and make a best faith effort.

12. **Licenses for existing assets**

Question: Are the creators or original owners of assets (e.g., code, data, models), used in the paper, properly credited and are the license and terms of use explicitly mentioned and properly respected?

Answer: [N/A]

Justification: The paper does not use existing assets.

Guidelines:

- The answer NA means that the paper does not use existing assets.
- The authors should cite the original paper that produced the code package or dataset.
- The authors should state which version of the asset is used and, if possible, include a URL.
- The name of the license (e.g., CC-BY 4.0) should be included for each asset.
- For scraped data from a particular source (e.g., website), the copyright and terms of service of that source should be provided.
- If assets are released, the license, copyright information, and terms of use in the package should be provided. For popular datasets, `paperswithcode.com/datasets` has curated licenses for some datasets. Their licensing guide can help determine the license of a dataset.

- For existing datasets that are re-packaged, both the original license and the license of the derived asset (if it has changed) should be provided.
- If this information is not available online, the authors are encouraged to reach out to the asset's creators.

13. **New Assets**

Question: Are new assets introduced in the paper well documented and is the documentation provided alongside the assets?

Answer: [Yes]

Justification: We introduce new assets in the form of evaluations of datasets using our framework. The process of performing these evaluations is discussed in Section 3.

Guidelines:

- The answer NA means that the paper does not release new assets.
- Researchers should communicate the details of the dataset/code/model as part of their submissions via structured templates. This includes details about training, license, limitations, etc.
- The paper should discuss whether and how consent was obtained from people whose asset is used.
- At submission time, remember to anonymize your assets (if applicable). You can either create an anonymized URL or include an anonymized zip file.

14. **Crowdsourcing and Research with Human Subjects**

Question: For crowdsourcing experiments and research with human subjects, does the paper include the full text of instructions given to participants and screenshots, if applicable, as well as details about compensation (if any)?

Answer: [N/A]

Justification: The paper does not involve crowdsourcing nor research with human subjects.

Guidelines:

- The answer NA means that the paper does not involve crowdsourcing nor research with human subjects.
- Including this information in the supplemental material is fine, but if the main contribution of the paper involves human subjects, then as much detail as possible should be included in the main paper.
- According to the NeurIPS Code of Ethics, workers involved in data collection, curation, or other labor should be paid at least the minimum wage in the country of the data collector.

15. **Institutional Review Board (IRB) Approvals or Equivalent for Research with Human Subjects**

Question: Does the paper describe potential risks incurred by study participants, whether such risks were disclosed to the subjects, and whether Institutional Review Board (IRB) approvals (or an equivalent approval/review based on the requirements of your country or institution) were obtained?

Answer: [N/A]

Justification: The paper does not involve crowdsourcing nor research with human subjects.

Guidelines:

- The answer NA means that the paper does not involve crowdsourcing nor research with human subjects.
- Depending on the country in which research is conducted, IRB approval (or equivalent) may be required for any human subjects research. If you obtained IRB approval, you should clearly state this in the paper.
- We recognize that the procedures for this may vary significantly between institutions and locations, and we expect authors to adhere to the NeurIPS Code of Ethics and the guidelines for their institution.
- For initial submissions, do not include any information that would break anonymity (if applicable), such as the institution conducting the review.

# The State of Data Curation at NeurIPS: Appendix

## A. Rubric

| | CURATORIAL ELEMENT | DESCRIPTION | DOCUMENTATION LEVEL | |
|---|---|---|---|---|
| | | | Criteria to meet minimum standard | Criteria to meet standard of excellence |
| SCOPE | | | | |
| 1 | Context, purpose, motivation | This information explains the purpose of dataset creation for the specified domain. | Documentation discusses the problem domain, what problems the new dataset addresses, the relevance of those problems, and the need for a new dataset in comparison to existing datasets. | Documentation explains how the context of the dataset affects possible reuse and includes reflection on the dataset creators' awareness of social, political, and historical context. |
| 2 | Requirements | The translation process from a "real-world" problem to a "ML problem" for which the dataset is created [98, 105] consists of numerous decisions, expertise, and worldviews that should be documented in order to understand the context in which the problem situation was framed. | Documentation states how the problem was formulated and how the dataset creation plan was generated. | Documentation includes reflection on how the problem formulation introduces intrinsic biases. |
| ETHICALITY AND REFLEXIVITY | | | | |
| 3 | Ethicality | Ethical considerations are critical to the fair and accountable creation and (re)use of datasets. | Documentation discusses how the benefits of creating the dataset outweigh any harms of creating it (see proportionality principle), and it discusses informed consent if the dataset is about humans. | Documentation goes beyond requirements listed in ethics framings like guidelines/policies/checklists. For example, documentation discusses alternate methods of dataset creation that were not used because of potential ethical harm. |
| 4 | Domain knowledge & data practices | Creating a dataset involves, often tacit, expertise about one or more domains as well as data practices. Articulating both types of nuance required in dataset development makes data work more transparent [42, 56, 98, 109, 135]. | Documentation states the domain-specific expertise and data skills required in developing the dataset. | Documentation discusses the required expertise needed to understand the intended purpose of the dataset and to reuse it. |
| 5 | Context awareness | Context awareness demonstrates an understanding of the subjective, non-neutral nature, and situatedness of data. | Documentation includes a positionality statement. | Documentation adopts a reflexive approach to dataset development. For example, documentation discusses how field epistemologies impact assumptions, methods, or framings. |

| 6 | Environmental footprint | This element is for dataset creators to reflect and quantify the footprint of their dataset creation process [6]. | Documentation contains a quantitative assessment of environmental footprint and clearly defined scope of what was measured. | Documentation includes a lifecycle assessment and the corresponding environmental footprint, and an assessment of design choices and rationale for the choices. |
|---|---|---|---|---|
| | | | DATA PIPELINE | |
| 7 | Data collection | Disclosing data sources is essential in the data collection process. Further reflection on the process of selecting those sources can reveal important interpretive assumptions [98] and historical and representational biases [56]. | If data was collected, documentation states how and why data and metadata were collected from the data source(s).<br><br>If data was synthesized, documentation discusses: 1) how and why the data was synthesized and 2) whether the data was synthesized to match labels, if used. | If data was collected, documentation discusses the process of defining criteria for selecting data source(s), specifies the criteria, explains why those criteria were chosen, and how the selected data sources are evaluated against these criteria.<br><br>If data was synthesized, documentation includes a reflection on potential intrinsic biases of the synthesis process, how the synthesis process shaped the features of the data, the limitations of the synthesis process, and how the synthesized data relates to the real-world distribution of the data it represents. |
| 8 | Data processing | Data processing involves cleaning, transforming, and wrangling data. Data processing decisions have impacts on the ultimate "cleaned" data that is used [77, 98]. Detailed documentation of this process enables outcomes of the model to be traced back to processing decisions. | Documentation discusses the process of cleaning, transforming, or wrangling data. | Documentation goes beyond what is done to discuss how the decisions about data processing were made and why, and potential impacts of the processing decisions. |
| 9 | Data annotation | Data annotation or labelling, regardless of the guidelines provided to reduce worker bias, can lead to disagreements on how data should be annotated (either between annotators or between dataset creators and annotators).The inclusion of this documentation highlights what is considered the "ground truth" [22, 98, 99] | Documentation discusses the process of annotation. If any labels are used, the documentation includes the following:<br><br>If labels are derived from the data: documentation discusses how data was interpreted to generate labels. | Documentation discusses the process of annotation with depth and reflexivity by including a reflection on how annotations (including labels, if used) represent differing worldviews and social backgrounds. |

| | | | | |
|---|---|---|---|---|
| | | by the dataset creators which impacts how annotation is performed [57]. | If the labels were created first and the data was derived from the labels: documentation discusses how the relationship of the data to the labels was verified.<br><br>If labels are obtained from elsewhere: documentation discusses where they were obtained from, how they were reused, and how the collected annotations and labels are combined with existing ones. | Additionally, if labels are derived from the data: documentation discusses how the labels are robust, i.e., not sensitive to variability and how disagreements on annotation were reconciled. |
| colspan=5 | DATA QUALITY |
| 10 | Suitability | Suitability is a measure of a dataset's quality with regards to the purpose defined. | Documentation discusses how the dataset is appropriate for the defined purpose. | Documentation discusses how dimensions such as accuracy, completeness, timeliness, and consistency contribute to the quality of the dataset in being used for the defined purpose. For example, timeliness (i.e., age) of data should be appropriate for the defined purpose. |
| 11 | Representativeness | Representativeness is a measure of how well a sample set of data represents the entire population. Sampling procedures and decisions about data sources can introduce extrinsic bias [98]. For example, choosing Reddit or Twitter as a data source can perpetuate dominant social biases rather than being a representative sample of the target population [6]. | Documentation defines the population and discusses the extent to which the sampling procedure is representative of the population. | Documentation includes reflection on how the dataset creation process overall, and the sampling procedures specifically, affect extrinsic bias. |
| 12 | Authenticity | Authenticity of a dataset is about whether the dataset "is what it purports to be" [23, 25, 26, 46, 110], which is a responsibility of dataset creators [74]. Authenticity can be established by assessing the identity and the integrity of the record [23, 24, 55, 69, 81, 84]. Integrity of a dataset is about whether "the material is complete and unaltered" [7, 13, 27, 46, 95]. | Documentation discusses how authenticity has been established and maintained, i.e.,<br>• Has the identity and origin of all data been verified?<br>• For data that is obtained, it is clear how the dataset creators have verified the identity of the dataset they reuse. | Documentation states how others can establish the authenticity of this dataset, i.e.,<br>• Documentation provides a persistent identifier and provenance information for the dataset in order for reusers to establish identity. |

| | | | | |
|---|---|---|---|---|
| | | | • For data that is generated, it is clear how they have been created and by whom.<br>• Has the integrity of all data been verified?<br>  • For data that is processed in any way, it is clear how processing steps may have impacted integrity. | • Documentation provides mechanisms for reusers to verify the integrity of their dataset. |
| 13 | Reliability | Reliability is about how well the dataset is "capable of standing for the facts to which it attests" [23], i.e., how certain we can be that its data points reflect what they represent. | Documentation discusses how the reliability of the dataset has been established and maintained, including the verification steps taken to ensure reliability, where necessary, i.e.,<br>• It is clear for each data element what synthetic or real-world phenomenon it represents. | Documentation states how others can establish the reliability of the dataset, i.e.,<br>• Documentation provides mechanisms to enable verification of what synthetic or real-world phenomenon each data element represents. |
| 14 | Structured documentation | Context documents in standardized structures provide information on the content of the dataset which is critical in establishing its usage in a well defined format. | Documentation includes a standardized context document. Acceptable formats include context documents that follow an established structure such as datasheets, data statements, and nutrition labels. | The context document addresses all mandatory items. |
| DATA MANAGEMENT | | | | |
| 15 | Findability | Ensuring findability is about enabling the dataset to be discovered for reuse after its development [138]. | Documentation discusses how the dataset is findable by providing a globally unique and persistent identifier (URLs are not persistent). | Documentation includes metadata and both the metadata and data are stored in a searchable repository. |
| 16 | Accessibility | Accessibility is about enabling the dataset to be obtained after its development [138]. | Documentation states all information and tools required to access the content of the data, and the identifier navigates to the metadata and data. | Documentation includes a communications protocol, an authentication and authorization procedure, and provides metadata that will be available even if data access is removed. |
| 17 | Interoperability | Interoperability ensures that the dataset can be integrated with other applications and workflows [138]. | Documentation discusses how the dataset integrates with other data, workflows, applications, etc. (i.e., that both the metadata and data are readable by humans and machines). | Documentation has metadata and data that both use controlled vocabularies and link to other resources using qualified references. |

| 18 | Reusability | Ensuring reusability requires providing information such as relevant provenance and usage [138]. | For both metadata and data, provenance information includes at least all of the following: 1) where the data came from, 2) who collected it, and 3) when it was collected. | Documentation has metadata and data that are both described using domain-relevant standards, state license and usage information, and provide additional provenance documentation as described by FAIR best practices. |
|----|-------------|--------------------------------------------------|---------------------------------------------|----------------------------------------|

## B. Rubric worksheet

| | CURATORIAL ELEMENT | DOCUMENTATION LEVEL | | | |
|---|---|---|---|---|---|
| | | Criteria to meet minimum standard | | Criteria to meet standard of excellence | |
| | | Pass/Fail | Comments | Full/Partial/None | Comments |
| | SCOPE | | | | |
| 1 | Context, purpose, motivation | | | | |
| 2 | Requirements | | | | |
| | ETHICALITY AND REFLEXIVITY | | | | |
| 3 | Ethicality | | | | |
| 4 | Domain knowledge & data practices | | | | |
| 5 | Context awareness | | | | |
| 6 | Environmental footprint | | | | |
| | DATA PIPELINE | | | | |
| 7 | Data collection | | | | |
| 8 | Data processing | | | | |
| 9 | Data annotation | | | | |
| | DATA QUALITY | | | | |
| 10 | Suitability | | | | |
| 11 | Representativeness | | | | |
| 12 | Authenticity | | | | |
| 13 | Reliability | | | | |
| 14 | Structured documentation | | | | |
| | DATA MANAGEMENT | | | | |
| 15 | Findability | | | | |
| 16 | Accessibility | | | | |
| 17 | Interoperability | | | | |
| 18 | Reusability | | | | |

## C. Toolkit

### C.1 Overview of research

**Background and Motivation**: The usage of artificial intelligence has increased exponentially with applications in predicting outcomes related to education, employment, housing, and many more social, economic, and financial aspects of our lives. Archival studies have long dealt with large amounts of data and concerns of representativeness, ethics, integrity, and more with the use of data curation methods, theories, and frameworks. Machine learning research (MLR) has pinpointed the data underlying predictive models to be the largest contributor in introducing bias [107, 117, 121]. Emerging studies have advocated for the prioritization of rigorous data curation practices often referred to as "data work" or "dataset development" in MLR [6, 42, 60]. Introducing data curation concepts and principles can therefore improve the transparency and accountability of the dataset creation process within MLR.

**Objectives:** We assess ML dataset development processes using principles and methods from archival studies and digital curation. We perform a synthesis and organization of existing work to enable the coherent usage of data curation frameworks, a taxonomy of data curation terms used within machine learning research, and a review of gaps and opportunities for data curation in machine learning.

**Method**: Our research design for this study consists of the following:
1. Synthesizing literature on data curation concepts and principles central to ML data work.
2. Exploring the relevance of data curation concepts and principles through an illustration of how they can be adapted, translated, and operationalized for ML data work.
3. Demonstrating the gaps and overlaps in how ML data practices already perform data curation, how data curation is discussed in MLR, and how data curation can be further adopted.

**Goals and contributions**: This project deepens the scholarly and practical connections between the data curation and machine learning research communities and initiate directions for improvement within MLR's data practices. The outcomes present a novel perspective on improving documentation practices in machine learning through data curation. Through this project, we aim to further establish the connections between the data curation and machine learning research communities.

### C.2 Application guidance

**Scope of application:** The rubric is intended for two types of users.
1. Firstly, dataset creators can use the rubric as a resource to prompt and facilitate critical engagement and reflection throughout their dataset creation process.
2. Secondly, existing datasets can be evaluated prior to publishing or reuse by applying the rubric to determine gaps that require further documentation and areas where bias can be introduced. In both cases, we aim for the rubric to be a practical and useful resource for researchers to engage with the dataset creation process using a data curation lens. The rubric was developed for the evaluation of ML datasets and has elements specific to the domain, including: requirements, data annotation, environmental footprint, and structured documentation.

### C.2.1 Applying the rubric to your own dataset

The overall process for using the rubric is as follows:

1. Read the rubric to get familiarized with the elements and details that will be needed.
2. Review each element in the rubric individually.
   a. For each element, first assess whether the minimum standard of documentation has been fulfilled. To do this, provide a pass/fail evaluation, where a pass is granted if all aspects specified under the minimum standard were discussed and a fail if they were only partially discussed or not discussed at all.
   b. Next, assess whether the documentation meets a standard of excellence, only if the minimum criteria received a pass. The standard of excellence is a full/partial/none evaluation. A full is

granted if all aspects specified in the standard of excellence column were discussed, a partial is granted if one or more (but not all) were discussed, and a fail if none were discussed.

    c. It is important to note both for points 2a and 2b that the quality of the responses/documentation is not being assessed but rather if the element was considered and reflected on in any capacity. The purpose of the rubric is to demonstrate the dataset creators' thought process and provide transparency so that its reuse is based on a complete understanding of the dataset.

3. For each element, along with the grade, a comment on what specific information was used to determine that grade must be provided. Other comments and questions can also be included.

The evaluation of each dataset can take 30-60 minutes.

## C.2.2 Applying the rubric to existing datasets through publications

The overall process for using the rubric is as follows:

1. Read the rubric to get familiarized with the elements and details that will be needed.
2. Gather and review all pertinent information that can be found about the dataset. This will include the research paper, appendices, the linked dataset, and any documentation associated with the externally linked dataset (e.g., README on github).
3. Review each element in the rubric individually by looking for it across all the information gathered in step 1. Some of the elements will be easier to locate than others because they will be titled specifically, whereas others may be discussed at any point.
    a. For each element, first assess whether the minimum standard of documentation has been fulfilled. To do this, provide a pass/fail evaluation, grant a pass if all aspects specified under the minimum standard were discussed and a fail if they were only partially discussed or not discussed at all.
    b. Next, assess whether the documentation meets a standard of excellence, only if the minimum criteria received a pass. The standard of excellence is a full/partial/none evaluation. A full is granted if all aspects specified in the standard of excellence column were discussed, a partial is granted if one or (but not all) were discussed, and a fail if none were discussed.
    c. It is important to note both for points 2a and 2b that the quality of the responses/documentation is not being assessed nor the correctness of the technicalities but rather if the element was considered and reflected on in any capacity. The purpose of the rubric is to demonstrate the dataset creators' thought process and provide transparency so that its reuse is based on a complete understanding of the dataset and how it was developed.
4. For each element, along with the grade, a comment on what specific information was used to determine that grade must be provided. Other comments and questions can also be included.
5. For each dataset, evaluators must provide a reflection on their overall assessment of the documentation and rigour demonstrated in the dataset creation process.
6. For each dataset, evaluators must provide a confidence rating for their evaluation.

We estimate the evaluation of each dataset will take about 30-60 minutes once you are familiar with the framework.

## C.2.3 How to interpret authenticity, reliability, and representativeness

It may be worth noting that the archival and digital curation perspectives that inform the evaluation framework are particularly important to interpreting the meaning of certain dimensions. Above all, the cluster of authenticity, integrity and reliability needs to be understood from this angle. They are closely related aspects, often treated or addressed by similar mechanisms, but they can be seen as analytically separate concepts. Here is an example.

When you download a data set of weather observations from a platform, you may want to verify if the file you have downloaded in fact is the data set you wanted to get, i.e., is it an authentic copy? You may be able to verify this with various checksums, both on the level of the file (e.g. a hashcode of the file, as commonly provided for downloads) and on the level of observations in some cases. In this case, you are concerned with **authenticity** - you want to verify that the data set is what it purports to be.

Authenticity does not guarantee you, however, that the observations in the data set are any good. A good observation of weather data is one that you can rely on to accurately represent how the weather actually was at the temporal and spatial locations covered by the data. Other aspects of goodness are reflected in the many quality standards for data, but when you want the data set to be able to stand in for the facts it represents, you are concerned with **reliability.** In other words, reliability is very much about the relationship of the data to whatever it represents. If the data set is a compilation of social media posts, then reliability will relate to the question whether these contributions were really posted, etc.

**Integrity** on the other hand refers to questions of tampering, errors, etc. For example, a dataset that lacks integrity is one for which we cannot assert that it contains *all* the items it originally contained, or that none of the items have been altered, falsified, or faked.

Consider a textbook case for records and archives for the difference between the three. A *passport* is a document that comes with very special features to prove that it can *stand in for the fact* that you are a citizen of the issuing country. Its **integrity** refers to the question whether it has been tampered with - has the photo been peeled off, have pages been removed or added? etc. The passport comes with features to prevent and check integrity. Its **authenticity** refers to the fact that it is indeed a passport of that country and that it indeed asserts the facts it states. Most of its special features are designed to make it easy to verify that (cf. banknotes). But imagine: a government could issue a perfectly authentic passport for a person who doesn't exist. That would be authentic, but it would not be reliable. The **reliability** rests on the relationship to the person it represents. We trust an authentic passport to be reliable because we trust the processes that governments have instituted and honed over the centuries to ensure that passports are only *issued to* authenticated citizens. But border control will use a machine-readable passport to look up and compare the information shown with the information stored in a database. When they do that, they verify reliability. For a deep dive into the archival perspective on what makes records authentic and reliable, see [5, 9].

Consider next a digital photograph taken during sunlight with a pro-grade digital lens reflex camera of a pantone color set of *whites* with standardized, specified colors, where the white balance is erroneously set at 'fluorescent light'. White balance relates to the color temperature of light: our eyes automatically adjust to different color temperatures, but a digital sensor does not. How an image looks on a screen is the result of computing it. In this case, the colors will not look very white on the photo without corrections to where the 'white point' should be located. The photo as taken is an **authentic** photo providing an **unreliable** representation of its subject. If you transfer the photograph yourself out of the camera you can also put in place mechanisms to verify integrity (including fixity checks and integrity checks using hash sums and the like on the file).

If you notice the error in color and then manually edit the binary code of the RAW file to set the white balance to the correct 'sunlight' setting, the photograph would in fact *lose* the property of 'integrity' since it has been tampered with (the hashcodes won't match), and it would *lose* the property of 'authenticity' since that was not the original setting, but it would gain in 'reliability' since the resulting color rendering would be a more accurate representation of how the colors should look. In this particular case, the fact that the subject of the photograph is standardized provides a *ground truth* that aids in verifying and assessing the photograph. Professional image processing software will be able to document both the 'as-taken' setting and 'to-use' setting of the photograph. Most photos, of course, are of subjects where this is much harder, and if the photograph is directly processed into a JPEG file, correcting white balance is much more difficult.

Finally, **representativeness** is related to reliability, but its perspective is much more narrowly focused on the question whether a data set accurately *represents* the overall set of observations or entities that it claims to be a sample of. For instance, for a data set of social media posts, the question will arise if it's representative of all platforms, all users, all topics, all media types, or various combinations of dimensions. All the statistical concepts around sampling apply as usual. Other data sets are not sampled out of an identified population but claim to stand for a general category so that representativeness is evaluated analytically, and so on.

### C.2.4 How to interpret findability, accessibility, interoperability, and reusability (FAIR)

Note that this group of criteria are a direct representation of the widely used FAIR principles [138] for research data sets, adopted and adapted for machine learning. We provide a simple checklist to assess whether the documentation of the dataset discusses the application of FAIR principles. This checklist is derived from the following tools and resources:

- Minglu Wang and Dany Savard. 2023. The FAIR Principles and Research Data Management. (September 2023). https://doi.org/10.5206/EXFO3999
- FAIR data maturity model
- https://zenodo.org/records/5111307#.Yj3Vi5rMI-Q
- https://ardc.edu.au/resource/fair-data-self-assessment-tool/
- https://fairaware.dans.knaw.nl/

1. Findable
   a. A globally unique (cannot be reused by someone else) and persistent (valid over time) ID (like DOI) is assigned to the data.
   b. The dataset is described by metadata (PID, license, description, provenance, etc.). Further guidelines and definitions of provenance can be found from the DCMI and our glossary.
   c. The metadata specifies the identifier.
   d. The metadata and data is stored in a searchable repository.
2. Accessible
   a. The identifier navigates to the metadata and data.
   b. Retrieval of the data is specified by a standard communications protocol (i.e., all information and tools that are required are communicated to access the content of the dataset) which is open and free to access.
   c. The communications protocol specifies the authentication and authorization procedure, if needed (i.e., if the dataset is not open and free-to-access, the protocol specifies how access would be granted).
   d. The metadata record is available even if the data is not.
3. Interoperable
   a. Metadata and data are *in principle* readable by humans and machines (i.e., has a structured format, open standard).
   b. Metadata and data use controlled vocabularies (standardized and universal terms for indexing and information retrieval). Metadata standards can be found in the RDA Metadata Standards Catalog (https://rdamsc.bath.ac.uk/).
   c. Metadata and data is linked to other metadata and data using qualified references (i.e., relationship to the resource is specified).
4. Reusable
   a. Metadata and data are well-described as per domain-relevant standards, have detailed provenance (where did the data come from, who collected it, when, etc.), and clear and accessible license and usage information.

### C.2.5 Guiding principles

We specify the following principles as "rules of thumb" to guide the evaluation of datasets:

1. Evaluate explicit documentation.

Evaluations should be made on the basis of documentation provided by the dataset creators, rather than performing evaluations ourselves.

2. Provide traceable comments.

The comments provided in the rubric to support the grade for each element should make recoverable the basis for the evaluation.

3. Minimum is easy, excellence is hard.

The evaluations for the minimum standard are meant to be *generous*. On the other hand, the standard of excellence criteria advocates for a high level of criticality, which is significantly harder to attain (compared to the minimum standard). The evaluations should only grant a 'Full' if all criteria are satisfied.

4. Don't make excuses.

If there is no documentation provided to evaluate an element, then don't make excuses for the dataset creators and evaluate it yourself or think of it as unnecessary. If you truly feel the element does not apply for that dataset, then that means it's feedback for the rubric and that the element needs further work so it applies to all types of datasets.

### C.2.6 Reflections & recommendations

In addition to the instructions on the process of using the rubric to evaluate datasets, the following recommendations are provided based on common reflections, challenges, and questions:

1. Completing an evaluation using the rubric requires iteration. A single pass through the rubric is often insufficient, especially for datasets that include various sources of documentation. The first iteration should be a step-by-step completion of each element in the rubric by looking for relevant information, keywords in the research paper or other dataset documentation. However, in doing so, sections of the documentation may be missed. It is therefore suggested to first evaluate the dataset by applying the rubric sequentially and then reviewing all the dataset documentation sequentially. The final step should be iterating as needed and zooming out.
2. The evaluation of elements will be interconnected, there can be notes to refer to the comment for another element.
3. If a context document is provided, it must be used to evaluate the elements. Although, the document will only provide information to fill in gaps rather than be sufficient to completely evaluate any element.
4. None of the elements should receive an N/A comment or grade.
5. The standard of excellence criteria should only be evaluated if the minimum standard criteria passes.
6. A failure for any element should not be provided based on the quality of the dataset but rather the documentation and reflection on the process of developing the dataset. For example, if the documentation acknowledges that the sample is not representative and can therefore introduce a bias- this is not considered a 'Fail'.
7. It is important to not evaluate the technical details provided but only evaluate the documentation. This means that evaluators should refrain from inferring the thought process or intention of the dataset creators based on their technical understanding of why the creators would develop their dataset in one way versus another. It is key to rely on explicit documentation only. This is important because the rubric assesses critical reflection around the dataset process not the quality of the dataset developed.

### C.2.7 FAQ

1. Is there a difference between labeling and annotation?

Please refer to the glossary for definitions differentiating the two terms. The rubric doesn't require evaluation of the "labeling" process if the dataset does not have labels.

2. How to evaluate consistency and timeliness for suitability?

Data quality is often defined as fitness for purpose and is multi-dimensional, meaning that it's measured through more than one data quality dimension such as accuracy, completeness, etc. Suitability, in the rubric, evaluates whether dataset creators ensure that their dataset's quality meets the purpose defined. For example, a dataset of math problems may not require timely data but may require consistent data (i.e., data presented in the same format). For standard of excellence, multiple data quality dimensions will apply for evaluation but potentially not all.

3. Is representativeness applicable to synthetic data?

Representativeness is still applicable to synthetic datasets because synthetic data is still representative of reality. However, this is a *conceptual* representativeness rather than a *statistical* one.

4. Why does the evaluation criteria for authenticity discuss data processing specifically?

Data processing alters the authenticity of a digital object. Authenticity is dependent on the bits of information in a file. For example, if you download a dataset with a hash code and make copies of it, all copies will have the same hash code. However, if you perform data processing (which changes the bits), the hash code will no longer be the same. In the rubric, for the minimum standard, you evaluate whether the dataset creators validate and verify the authenticity of the data they are collecting. Whereas for standard of excellence, you evaluate whether they have

processes to ensure people that reuse their dataset are able to claim authenticity (i.e., maintaining the chain of authenticity).

5. For the data quality elements, are we evaluating that the dataset is suitable, authentic, has integrity, is representative, and is reliable OR that the dataset creators discuss their processes for ensuring these? If there is no mention of these qualities specifically, how do we evaluate them?

For data quality elements, you are evaluating whether the dataset creators discussed their processes for ensuring that their dataset is suitable, authentic, reliable, has integrity, and the extent to which it is representative (and why if it is not). Remember the guiding principle- "evaluate explicit documentation". We have added another guiding principle- "don't make excuses". If no documentation is provided for these data quality elements, then don't make excuses for the dataset creators and evaluate it yourself or think of it as unnecessary. If you truly feel the element does not apply for that dataset, then that means it's feedback for the rubric and that the element needs further work so it applies to all types of datasets.

6. Does hosting a dataset on huggingface make it 'findable'?

It depends, if it's hosted on huggingface but does not have a persistent identifier like a DOI, then it is not findable. See next question.

7. Why are URLs not acceptable for findability?

URLs are not considered "findable" because of the high likelihood of link rot (that the link over time will no longer be available). There are studies that show that academic papers are highly perceptible to link rot, eg: see [62]. Instead, we want persistent identifiers like DOIs to make sure the dataset is findable in the future.

8. What is the difference between findability and accessibility?

Findability is about a dataset being easily located. For example, if a publication provides a zenodo link to a dataset, that would make it findable (zenodo assigns a DOI to everything it publishes). So here we're looking for a dataset being easily located, indexed, catalogued, etc.

Accessibility is about whether a dataset can be opened and used and read. For example, is it in a format you can read, can you download it (i.e., is it retrievable), is the access blocked off via password-protection, are there access and authorization protocols?

A dataset would then be findable if there was a link pointing to it but not accessible if you couldn't open it because you didn't have the password for it and there was no documentation of an access protocol. On the other hand, if a dataset was open-access (eg, through github) but didn't have a persistent identifier (eg DOI) and wasn't indexed in a repository like zenodo then it would be accessible but not findable. Since accessibility rests on *accessing* the content, a URL alone is not enough to make it accessible either. So even if the dataset is available through github there must be other documentation that provides any further information needed to access the content and metadata.

9. Can you provide further clarification for evaluating interoperability (especially standard of excellence)?

For the minimum standard, the documentation must explain how the dataset can be integrated with other data and workflows. An example of that is that the data can be exported to popular, standard formats. For the standard of excellence, the data and metadata must use controlled vocabularies and link to other resources with qualified references. For example, metadata can be created using controlled vocabularies like the W3C's Data Catalog Vocabulaire (DCAT) model which defines terms like dataset vs data service, catalog (as a subclass of dataset), and so on. Please see this blurb from FAIR about qualified references:

> "A qualified reference is a cross-reference that explains its intent. For example, *X is regulator of Y is a much more qualified reference than *X is associated with Y*, or *X see also Y*. The goal therefore is to create as many meaningful links as possible between (meta)data resources to enrich the contextual knowledge about the data, balanced against the time/energy involved in making a good data model. To be more concrete, you should specify if one dataset builds on another data set, if additional datasets are needed to complete the data, or if complementary information is stored in a different dataset. In particular, the scientific links between the datasets need to be described. Furthermore, all datasets need to be properly cited (i.e., including their globally unique and persistent identifiers)." [33].

Zenodo also has a [webpage](#) that describes how it fulfills the FAIR principles for its datasets [143].

**C.3 Rubric**

See [Section A](#) of the Appendix.

**C.4 Rubric worksheet**

See [Section B](#) of the Appendix.

**C.5 Sample evaluations**

Please note that the sample evaluations were performed using the version of the rubric at the time of evaluating datasets from round 3. Note also that the description column and cited references are deleted below for space, see full [rubric](#) with references.

Paper: FS-Mol: A Few-Shot Learning Dataset of Molecules [129]

| | CURATORIAL ELEMENT | DOCUMENTATION LEVEL | | | |
|---|---|---|---|---|---|
| | | Criteria to meet minimum standard | PASS/ FAIL | Criteria to meet standard of excellence | Full/ Partial/ None |
| | | | SCOPE | | |
| 1 | Context, purpose, motivation | Pass | Paper introduction discusses the problem domain and why a new dataset is needed; see 'related work' in paper and appendix B in supplementary material ('related work details') for comparison to existing datasets. | Full | Section 7 of paper discusses how dataset can be used outside of its original context ("it is now possible to evaluate… we note that transfer of results to realistic projects is not guaranteed to be successful…") |
| 2 | Requirements | Pass | Section 2 of paper (especially " 2.2 Desired Attributes of a QSAR Few-Shot Dataset and Benchmark") explicitly derives design requirements to create the dataset. | Partial | No explicit discussion of intrinsic biases introduced by problem formulation; other approaches to formulating the problem are discussed in 'related work' section of paper (discussing other datasets and their features) |
| | | | ETHICALITY AND REFLEXIVITY | | |
| 3 | Ethicality | Pass | No discussion of consent (no human data); pg 9 'societal impacts' section discusses benefits of creating the dataset. | Fail | No additional discussion of ethical consideration throughout the paper or supplementary documentation. |
| 4 | Domain knowledge & data practices | Pass | On pg 2 of papers, authors state aim to "demonstrate the utility of few-shot learning methods in an important domain, namely QSAR, which does not provide an obvious generic pretraining corpus (such as in NLP or computer vision). The proposed dataset is specifically designed to replicate the challenges of machine learning in the very low data regime of drug-discovery projects" (focus on drug-discovery domain) | Partial | README in GitHub repo discusses activities to be undertaken to re-use the dataset "Hence, in order to be able to run MAT, one has to clone our repository via…" – not directly discussing any domain knowledge needed. |
| 5 | Context awareness | Fail | Research goals are described but not positioned relative to researchers' intellectual/political believes; researcher positions not disclosed/no positionality statement included. | None | Failed minimum criteria. |

| 6 | Environmental footprint | Fail | No assessment of environmental footprint | None | Failed minimum criteria |
|---|---|---|---|---|---|
| | | | DATA PIPELINE | | |
| 7 | Data collection | Pass | ExtractDataset.ipynb from GitHub repo describes how data were gathered by querying ChEMBL; section 3 of paper explains data acquisition process in detail ("the reason why we remove large assays is…") | Partial | Section B of supplementary material describes other few-shot learning and molecular property datasets (e.g. why they used ChEMBL instead of other sources); no explicit discussion of criteria for source selection, why criteria were chosen, or how other sources were validated against criteria. |
| 8 | Data processing | Pass | ExtractDataset.ipynb from GitHub repo describes how data were cleaned and split into test vs validation assays. | Full | Section 3 of paper describes decisions behind data processing (e.g. "In this way, our proposed meta-testing tasks closely mimic the new-lead optimization problem, where a completely unseen task is presented for adaptation.") |
| 9 | Data annotation | Pass | "Binary Classification Task" section of paper discusses some annotation activity | None | No discussion of robustness of annotations. |
| | | | DATA QUALITY | | |
| 10 | Suitability | Pass | Section 6 and first paragraph of section 7 describe and demonstrate dataset appropriateness for purpose. | Partial | Documentation does not explicitly discuss accuracy/completeness/timeliness of the chosen dataset, but Section 6 of the paper demonstrates the utility of the dataset for its intended purpose by providing "a set of results for all three categories of few-shot learning, with representative methods of the use of this dataset in each". |
| 11 | Representativeness | Pass | Section 3 on pg 3 of main paper describes how the 'sample' of the dataset is taken from the overall population (the ChEMBL database); also on pg 9 "the few-shot baselines we provide checkpoints and results for are only a representative set, rather than a complete survey of the current state of the field" | None | No explicit discussion of biases. |
| 12 | Authenticity | Pass | No explicit discussion of authenticity but extractdataset.ipynb does discuss how initial raw data were obtained (e.g. describes process by which database was queried) | Partial | No explicit discussion of future authenticity/preservation processes, but does discuss in section A of supp material how dataset documentation facilitates re-use more generally. |
| 13 | Reliability | Pass | Section 5 of paper discussing benchmarking procedures (i.e. making sure that the dataset is useful for what it's supposed to be useful for) | Partial | No explicit discussion of reliability management in the context of future re-use; section A of supplementary material discusses how the dataset documentation facilitates re-use. |
| 14 | Integrity | Fail | No discussion of dataset integrity or preservation processes (section H of | None | Failed minimum criteria. |

| | | | supplementary document does not actually discuss a maintenance plan or means of maintaining accuracy/consistency over time). | | |
|---|---|---|---|---|---|
| 15 | Structured documentation | Fail | No standardized context document | None | Failed minimum criteria |
| | | | DATA MANAGEMENT | | |
| 16 | Findability | Fail | No persistent identifier provided. | None | Failed minimum criteria |
| 17 | Accessibility | Pass | Section F of supplementary material describes computational resources used; GitHub README states the tools and steps required to access data content. | Partial | GitHub repo includes a code of conduct document, as well as protocols for contributing and for security reporting. |
| 18 | Interoperability | Pass | README in GitHub repo describes how to use the dataset with "three key few-shot learning methods"; dataset.ipynb describes the machine/human readable metadata. | Full | Dataset.ipynb describes the controlled vocabularies for specific dataclasses (e.g. task_name as a string describing the task each point is taken from) |
| 19 | Reusability | Fail | From data contents of GitHub repo it does not appear that data or metadata contain provenance information about where the dataset came from/when/who collected it; license is included in the GitHub repo. | None | Failed minimum criteria. |

Paper: American Stories: A Large-Scale Structured Text Dataset of Historical U.S. Newspapers [18]

| | CURATORIAL ELEMENT | DOCUMENTATION LEVEL | | | |
|---|---|---|---|---|---|
| | | Criteria to meet minimum standard | PASS/ FAIL | Criteria to meet standard of excellence | Full/ Partial/ None |
| | | | SCOPE | | |
| 1 | Context, purpose, motivation | Pass | Paper Introduction and section 6 (Applications) discusses the problems and relevance, and 'Related Literature' (section 2) discusses other similar datasets. | Full | 'Applications' section on pg 6 of supplementary material discusses "multiple applications that can be facilitated by the American Stories dataset" |
| 2 | Requirements | Pass | Paper Introduction (pg 2, "To address these limitations, we develop…") introduces certain requirements. | Partial | On pg 3 of paper ,documentation reflects on the bias potentially introduced by scanning illegible newspapers; other approaches are discussed in Section 2 on Related Literature (but not specifically other approaches the authors considered) |
| | | | ETHICALITY AND REFLEXIVITY | | |
| 3 | Ethicality | Pass | Some harms (e.g. offensive language) are discussed in Section 7: Conclusion. Consent is discussed in datasheet (pg 14 of supplementary material) | Partial | Some additional discussion of copyrights/accessibility on pg 3 of paper |
| 4 | Domain knowledge & data practices | Pass | Pg. 23 of paper (the datasheet) addresses the professors, research assistants, and students involved in data collection | Partial | Datasheet states "There are a large number of potential uses in the social sciences, digital humanities, and deep learning research" |
| 5 | Context awareness | Pass | No positionality statement but several mentions throughout the datasheet showing awareness of social context ("This dataset contains unfiltered content composed by newspaper editors, columnists, and other sources. It reflects their biases and any factual errors that they made."), and section 7 of the paper reflects on the historicity of dataset contents | Partial | Section 3 of paper touches on assumptions going into methodological choices (e.g. on pg 3, "We do not OCR ads because…") |
| 6 | Environmental footprint | Fail | No environmental assessment. | None | Failed minimum criteria |
| | | | DATA PIPELINE | | |
| 7 | Data collection | Pass | Described in 'Composition' (pg 11) and 'Collection Process' (pg 13) sections of datasheet in supplementary material | Partial | We have a lot of information about how the data were collected, but I still don't see where in the documentation it specifies the criteria they used to select data sources or how data sources were validated against these criteria (e.g. why the library of congress dataset?). |
| 8 | Data processing | Pass | Pre-processing section of datasheet (pg 14 of supplementary material) describes process of cleaning and wrangling data | Full | Sections 3, 4, and 5 of main paper discuss the implications of processing decisions (e.g. on computing cost and efficiency) |

| 9 | Data annotation | Pass | Student annotation is discussed ins Section 5 'Pipeline Evaluation' of main paper | Full | Student annotations were used as 'ground truth' for model training; see pg 5 of supplementary material |
|---|---|---|---|---|---|
| DATA QUALITY | | | | | |
| 10 | Suitability | Pass | Section 5 of paper evaluates the pipeline for accuracy, legibility, and comparison to other OCR engines | Full | See explanation for minimum criteria |
| 11 | Representativeness | Pass | Sampling approach discussed in datasheet (pg 13 of supplementary material) – it includes everything in the Chronicling American scan collection. | Full | Section 3 of paper discusses how illegible papers and their inclusion/exclusion in the dataset could bias results. |
| 12 | Authenticity | Pass | Pipeline for generating data is included in the Github repo (https://github.com/dell-research-harvard/AmericanStories?tab=readme-ov-file); no explicit discussion of authenticity | None | No explicit discussion of authenticity in future re-use. |
| 13 | Reliability | Pass | Section 5 of paper (Pipeline Evaluation) describes verification and validation processes used to ensure reliability. | Full | Maintenance section of datasheet discusses how errors will be corrected in future (and uploaded to HuggingFace) |
| 14 | Integrity | Pass | Documentation does not explicitly discuss integrity but datasheet does emphasize that "material is complete and unaltered" | Full | Maintenance section of datasheet describes preservation processes in place (e.g. old versions still accessible via HuggingFace) |
| 15 | Structured documentation | Pass | Paper and supplementary material include a datasheet (Gebru et al) | Full | All mandatory components of datasheet are answered. |
| DATA MANAGEMENT | | | | | |
| 16 | Findability | Pass | DOI available on HuggingFace page (10.57967/hf/0757) | Full | Data and metadata stored in searchable repo (HuggingFace) |
| 17 | Accessibility | Pass | Steps for accessing data listed on HuggingFace page data card and described in 'Distribution' section of datasheet (pg 15 of supplementary material | Full | Communications protocol described in 'Maintenance' section of datasheet (supp material pg 16) |
| 18 | Interoperability | Pass | Pg 4 of paper describes readable formats of metadata and data ("The raw files are in a json format, and the Hugging Face repo comes with a setup script that easily allows people to download both raw and parsed data to facilitate language modeling and computational social science applications."; lots of metadata info included on HuggingFace page | Full | See HuggingFace page for controlled metadata vocabularies |
| 19 | Reusability | Pass | Some provenance information included in metadata (e.g. where it came from, associated newspaper, but not who collected it/when) | Partial | Pg 16 of supplementary material (datasheet) states "The dataset is distributed under a Creative Commons CC-BY license. The terms of this license can be viewed at https://creativecommons.org/licenses/by/2.0/" |

## C.6 Glossary

Table 1: Glossary Terms

| Term | Definition | Discussion/Example | Sources |
|------|-----------|-------------------|---------|
| Context document | "Interventions designed to accompany a dataset or ML model, allowing builders to communicate with users". | Context documents are standardized documentation formats that convey information about the dataset, types of context documents include datasheets, nutrition labels, etc. | [11] |
| Data annotation | Although data annotation and labelling are often used interchangeably in ML, labelling is a subset of annotation. (See labelling)<br><br>Data annotation refers to the process of adding information to a dataset to provide more context. For example, adding metadata. | Annotation can include metadata about the units of measurement. | |
| Data practices | "What and how data are collected, managed, used, interpreted, released, reused, deposited, curated, and so on…" | Data practices are the decisions made in the collecting, interpretation, etc. of data. | [10] |
| Extrinsic bias | Extrinsic bias refers to bias that exists within the dataset which are reflections of social, historical biases. | "Extrinsic bias is concerned with a view of a biased dataset "from the outside." The argument is that an already-biased dataset can cause even innocent software to produce a biased outcome - and may look like people saying things such as "the data made me do it." … If we fail to remember that a dataset is biased, then we may treat it as "fair" or "representative," harming people who have been excluded from it." | [98] |
| Informed consent | Informed consent is a standard ethical principle of research with human subjects that rests on the commitment that participants<br>• Are fully informed<br>• Decide voluntarily<br>• Before research is conducted.<br>Its application in online environments is complicated by the shift in technology and methods (see [28]), but the principle remains important. | In conventional human subject studies such as interviews, an IRB reviews ethics protocols and evaluates if the research is compliant with principles such as the proportionality principle. In social media research, things get complicated. In some situations, implied consent (see [53]) can be present but must be justified. In the case of LLMS, widespread data collection without consent has prompted massive ethical and legal concerns. | A discussion of Twitter research ethics [28] is a good start. |
| Intrinsic bias | "The ways in which we change the data "from the inside" of data science work-processes while we are preparing the data for modeling."<br><br>Intrinsic bias is the bias data workers introduce to the dataset. | "Through practices of data wrangling, curation, and feature-engineering, humans make a series of decisions about how to treat their data, and those decisions may inadvertently introduce bias into the data." | [98] |

| Labelling | Labelling is a specific type of annotation that involves assigning a predefined category to a data item. | Labelling tweets on Twitter as 'human-generated' or 'bot-generated'. | |
|---|---|---|---|
| PID: persistent identifier | "Globally unique and persistent identifiers remove ambiguity in the meaning of your published data by assigning a unique identifier to every element of metadata and every concept/measurement in your dataset. [IDs] must be persistent. It takes time and money to keep web links active, so links tend to become invalid over time. Registry services guarantee resolvability of that link into the future, at least to some degree." | ORCID iDs are persistent identifiers for people. DOIs are persistent identifiers for journal articles, datasets, etc. | [34] |
| Population | Mathematical term used to describe a group of units sharing a common trait. | | |
| Positionality statement | "Researcher/Practitioner Self-Disclosure: Practice should involve a disclosure of the researcher's position in the world, her or his goals, as well as the researcher's position in her or his intellectual and, to an appropriate extent, political beliefs" | See [72] for examples. | [3] |
| Problem formulation | The process with which a problem is formulated and the methods we use to define and measure it define the lens with which we see the problem. "...the problems we solve with data science are never insulated from the larger process of getting data science to return actionable results…, these ends are very much an artifact of a contingent process of arriving at a successful formulation of the problem, and they cannot be easily decoupled from the process at arriving at these ends."[105:9]. | For example, O'Neil describes how proxies are used to quantify university excellence through indicators that are easily collected such as SAT scores, student-teacher ratios, and acceptance rates rather than through students' learning experience, happiness, productivity, personal fulfillment, etc. [101]. <br><br> See [105] for additional examples. | [105] |
| Proportionality principle | In ethics, it is understood that actions have positive and negative effects simultaneously. This is called *double effect*. "Applications of double effect always presuppose that some kind of proportionality condition has been satisfied. Traditional formulations of the proportionality condition require that the value of promoting the good end outweigh the disvalue of the harmful side effect." | In medicine, a surgeon may cause harm to a patient's skin (negative) in order to save their heart (positive). It would not be permissible for a surgeon to open up someone's chest just to get a better look or take a selfie with it because that would violate the proportionality principle. | [92] |
| Provenance | Provenance information provides a trail of history about how the data originated, how it's changed, who was involved, and more. | See the following blurb from the FAIR principles… <br><br> "For others to reuse your data, they should know where the data came from … who to cite and/or how you | [35] |

| | | |
|---|---|---|
| | wish to be acknowledged. Include a description of the workflow that led to your data: Who generated or collected it? How has it been processed? Has it been published before? Does it contain data from someone else that you may have transformed or completed?" | |
| Reflexivity | "Questions of reflexivity ask us to consider who we should listen to and why, how to place actors' ideas in a larger field of power, questions about our own relationship to actors' theories of the world. Reflexivity asks us to approach our work with epistemological unease because we are always at risk of reproducing categories that reify power." | [93, 132] |

## C.7 Further readings

The following readings 1) showcase how data curation is discussed in data science and machine learning studies, 2) contain context for relevant data curation terms, concepts, and frameworks, and 3) provide important terminology for ML benchmarks.

### C.7.1 Data curation in data science

A vast amount of literature points to the datasets used for training machine learning models to be the source for introducing bias in model results leading to a call for increased documentation of datasets used in ML. Emerging research has proposed context documents – "interventions designed to accompany a dataset or ML model, allowing builders to communicate with users" [11]. The following are types of relevant context documents.

1. Timnit Gebru, Jamie Morgenstern, Briana Vecchione, Jennifer Wortman Vaughan, Hanna Wallach, Hal Daumé III, and Kate Crawford. 2021. Datasheets for datasets. *Commun. ACM* 64, 12 (November 2021), 86–92. https://doi.org/10.1145/3458723

Datasheets are one of the most popular methods of documenting the process of developing datasets as well as providing a dataset description. This paper is a good introduction to how dataset documentation is evaluated [32].

2. Michael A. Madaio, Luke Stark, Jennifer Wortman Vaughan, and Hanna Wallach. 2020. Co-Designing Checklists to Understand Organizational Challenges and Opportunities around Fairness in AI. In *Proceedings of the 2020 CHI Conference on Human Factors in Computing Systems*, April 21, 2020, Honolulu HI USA. ACM, Honolulu HI USA, 1–14. https://doi.org/10.1145/3313831.3376445

Madiao et al. developed a resource - checklist for AI fairness - based on findings of current practitioners processes, needs, and requirements for developing fair AI models [85].

3. Emily M. Bender and Batya Friedman. 2018. Data Statements for Natural Language Processing: Toward Mitigating System Bias and Enabling Better Science. *Transactions of the Association for Computational Linguistics* 6, (2018), 587–604. https://doi.org/10.1162/tacl_a_00041

Bender and Friedman develop 'data statements'- a resource for NLP training datasets to be documented in order to mitigate bias and exclusion [5].

Topics like dataset documentation in ML are often discussed as a part of data practices, data work, or dataset development. The following studies talk about stages of dataset development processes, how data scientists or data workers approach their data work, and the importance and impact of decisions made during the dataset development.

1. Morgan Klaus Scheuerman, Alex Hanna, and Emily Denton. 2021. Do Datasets Have Politics? Disciplinary Values in Computer Vision Dataset Development. *Proc. ACM Hum.-Comput. Interact.* 5, CSCW2 (October 2021), 1–37. https://doi.org/10.1145/3476058

This paper discusses how documentation captures underlying values of data practices in machine learning (specifically computer vision tasks) [121]. Specifically, publications are analyzed to understand the documentation and communication of datasets. The findings showcase the practices that are silenced (such as data work, context, positionality, and care) over those that are (wrongly) embraced such as model work, universality, and so on. This reading helps reflect on and understand how intrinsic bias can be introduced within datasets.

2. Nithya Sambasivan, Shivani Kapania, Hannah Highfill, Diana Akrong, Praveen Paritosh, and Lora M Aroyo. 2021. "Everyone wants to do the model work, not the data work": Data Cascades in High-Stakes AI. In *Proceedings of the 2021 CHI Conference on Human Factors in Computing Systems*, May 06, 2021, Yokohama Japan. ACM, Yokohama Japan, 1–15. https://doi.org/10.1145/3411764.3445518

Through interviews with AI practitioners, Sambasivan et al. find that poor data practices in high-stakes AI domains (i.e., practices that do not prioritize data quality) lead to data cascades which are negative impacts of data issues [117].

3. Milagros Miceli, Julian Posada, and Tianling Yang. 2022. Studying Up Machine Learning Data: Why Talk About Bias When We Mean Power? *Proc. ACM Hum.-Comput. Interact.* 6, GROUP (January 2022), 1–14. https://doi.org/10.1145/3492853

Miceli et al. discuss that while we often recognize that there is bias in the datasets and their processes used for ML models, it is often ignored that this bias is a result of power inequities [93]. The authors analyze data bias, data work, and data documentation from a "power-aware" framing as compared to a "bias-oriented" one. This paper provides an interesting shift in perspective which further illuminates the importance of reflexivity in data work.

4. Michael Muller and Angelika Strohmayer. 2022. Forgetting Practices in the Data Sciences. In *CHI Conference on Human Factors in Computing Systems*, April 29, 2022, New Orleans LA USA. ACM, New Orleans LA USA, 1–19. https://doi.org/10.1145/3491102.3517644

This paper studies how data processing leads to different types of forgetting and where and how each type of forgetting occurs in the machine learning stack [98]. Forgetting is conceptualized as the practice that occurs when choices are made about what data is kept, what it represents and so forth (therefore by designing a dataset in a given way, we *remember* only its current state, and *forget* the decisions, the erased data, etc.). This is a great paper for a deep dive into the various types of design decisions that impact the eventual dataset.

The previous studies discuss aspects of data curation as dataset development. However, some ML studies have started discussing the importance of data curation by referencing archival studies and digital curation directly. These are included below.

1. Susan Leavy, Eugenia Siapera, and Barry O'Sullivan. 2021. Ethical Data Curation for AI: An Approach based on Feminist Epistemology and Critical Theories of Race. In *Proc. of 2021 AAAI/ACM Conf. on AI, Ethics, and Society*, July 21, 2021, Virtual Event USA. ACM, Virtual Event USA, 695–703. Retrieved November 11, 2022 from https://dl.acm.org/doi/10.1145/3461702.3462598

This study discusses principles for ethical data curation based on race critical race theory and data feminism to improve the reflection of power, bias, and values in data processes and thereby improve transparency and accountability of AI systems [68].

2. Emily M. Bender, Timnit Gebru, Angelina McMillan-Major, and Shmargaret Shmitchell. 2021. On the Dangers of Stochastic Parrots: Can Language Models Be Too Big? 🦜 . In *Proceedings of the 2021 ACM Conference on Fairness, Accountability, and Transparency* (*FAccT '21*), March 01, 2021, New York, NY, USA. Association for Computing Machinery, New York, NY, USA, 610–623. . https://doi.org/10.1145/3442188.3445922

This paper discusses the potential risks of language models (and by extension other ML/AI systems) [6]. The authors recommend a shift towards careful, reflective practices around datasets and model development along with a greater focus towards documentation.

3.  Eun Seo Jo and Timnit Gebru. 2020. Lessons from archives: strategies for collecting sociocultural data in machine learning. In *Proceedings of the 2020 Conference on Fairness, Accountability, and Transparency*, January 27, 2020, Barcelona Spain. ACM, Barcelona Spain, 306–316. https://doi.org/10.1145/3351095.3372829

This paper highlights that practices from archival studies have experience dealing with consent, power dynamics, transparency, and ethics and that these practices should be adopted into data collection and annotation practices in machine learning [56].

### C.7.2 Data curation

Data curation involves "maintaining and adding value to digital research data for current and future use" [20]. The following studies introduce data/digital curation terminology and the data curation lifecycle model (parallel to ML model pipelines) with the aim to familiarize how the data curation field approaches data work.

1.  Sarah Higgins. 2008. The DCC Curation Lifecycle Model. *International Journal of Digital Curation* 3, 1 (August 2008), 134–140. https://doi.org/10.2218/ijdc.v3i1.48

The paper introduces the curation lifecycle model by emphasizing it as a lifecycle (as opposed to a linear process). Each stage of the model is briefly introduced [45].

2.  Sarah Higgins. 2012. The lifecycle of data management. In *Managing Research Data* (1st ed.), Graham Pryor (ed.). Facet, 17–46. https://doi.org/10.29085/9781856048910.003

This paper discusses each stage in depth including the tasks performed, how each stage leads to the next, and the expected outcomes [47].

3.  Digital Curation Centre. Glossary. *Digital Curation Centre*. Retrieved January 21, 2024 from https://www.dcc.ac.uk/about/digital-curation/glossary

This is a glossary of common digital curation terms - to be returned to as a resource, as needed [21].

4.  Carole L Palmer, Nicholas M Weber, Trevor Muñoz, and Allen H Renear. Foundations of Data Curation: The Pedagogy and Practice of "Purposeful Work" with Research Data. 16.

This is an introductory paper to the field of data curation and its place within archival studies, library studies, and computer science [103].

### C.7.3 Benchmarking in ML

Benchmarking is often not a well discussed topic in machine learning papers. The below list is compiled to introduce commonly used terms including: benchmark dataset, benchmark tasks, simulator, synthetic dataset, baseline method, benchmark suite, etc.

1.  Matthew Stewart. 2023. The Olympics of AI: Benchmarking Machine Learning Systems. *Medium*. Retrieved January 21, 2024 from https://towardsdatascience.com/the-olympics-of-ai-benchmarking-machine-learning-systems-c4b2051fbd2b

This paper explains terms benchmark, benchmark dataset, benchmark tasks, baseline method, and benchmark suite [89].

2.  Ramona Leenings, Nils R. Winter, Udo Dannlowski, and Tim Hahn. 2022. Recommendations for machine learning benchmarks in neuroimaging. *NeuroImage* 257, (August 2022), 119298. https://doi.org/10.1016/j.neuroimage.2022.119298

This paper explains benchmark term and concept [71].

3.  Kim Martineau. 2021. What is synthetic data? *IBM Research Blog*. Retrieved January 21, 2024 from https://research.ibm.com/blog/what-is-synthetic-data

This paper explains the term *synthetic data* [61].

4.  Nataniel Ruiz. 2019. Learning to Simulate. *Medium*. Retrieved January 21, 2024 from https://towardsdatascience.com/learning-to-simulate-c53d8b393a56

This paper explains the term *simulator* [115].

## D. Additional information about rubric evaluations

### D.1 List of datasets

Table 2: List of datasets evaluated from the NeurIPS Datasets & Benchmarks track

| Dataset Number | Round | Paper Title | Dataset Abbreviation | Publication Year | Reference |
|---|---|---|---|---|---|
| 1 | Training | Programming Puzzles | program_puzzles | 2021 | [123] |
| 2 | Training | Open Bandit Dataset and Pipeline: Towards Realistic and Reproducible Off-Policy Evaluation | open_bandit | 2021 | [116] |
| 3 | Training | SciGen: a Dataset for Reasoning-Aware Text Generation from Scientific Tables | scigen | 2021 | [96] |
| 4 | Training | MOMA-LRG: Language-Refined Graphs for Multi-Object Multi-Actor Activity Parsing | moma-lrg | 2022 | [83] |
| 5 | Training | CEDe: A collection of expert-curated datasets with atom-level entity annotations for Optical Chemical Structure Recognition | cede | 2022 | [49] |
| 6 | Round 1 | LoveDA: A Remote Sensing Land-Cover Dataset for Domain Adaptive Semantic Segmentation | loveda | 2021 | [137] |
| 7 | Round 1 | RELLISUR: A Real Low-Light Image Super-Resolution Dataset | rellisur | 2021 | [1] |
| 8 | Round 1 | Measuring Mathematical Problem Solving With the MATH Dataset | math | 2021 | [44] |
| 9 | Round 1 | DGraph: A Large-Scale Financial Dataset for Graph Anomaly Detection | dgraph | 2022 | [142] |
| 10 | Round 1 | Change Event Dataset for Discovery from Spatio-temporal Remote Sensing Imagery | change_event | 2022 | [87] |
| 11 | Round 1 | CAESAR: An Embodied Simulator for Generating Multimodal Referring Expression Datasets | caesar | 2022 | [54] |
| 12 | Round 1 | GLOBEM Dataset: Multi-Year Datasets for Longitudinal Human Behavior Modeling Generalization | globem | 2022 | [141] |
| 13 | Round 1 | ClimateSet: A Large-Scale Climate Model Dataset for Machine Learning | climateset | 2023 | [58] |
| 14 | Round 1 | BubbleML: A Multiphase Multiphysics Dataset and Benchmarks for Machine Learning | bubbleML | 2023 | [41] |
| 15 | Round 1 | DataComp: In search of the next generation of multimodal datasets | datacomp | 2023 | [30] |
| 16 | Round 2 | The CPD Data Set: Personnel, Use of Force, and Complaints in the Chicago Police Department | cpd | 2021 | [48] |
| 17 | Round 2 | The Tufts fNIRS Mental Workload Dataset & Benchmark for Brain-Computer Interfaces that Generalize | tufts | 2021 | [51] |
| 18 | Round 2 | How Would The Viewer Feel? Estimating Wellbeing From Video Scenarios | viewer | 2022 | [90] |
| 19 | Round 2 | The RefinedWeb Dataset for Falcon LLM: Outperforming Curated Corpora with Web Data Only | refinedweb | 2023 | [108] |

| 20 | Round 2 | Stanford-ORB: A Real-World 3D Object Inverse Rendering Benchmark | stanfordorb | 2023 | [63] |
|----|---------|------------------------------------------------------------------|-------------|------|------|
| 21 | Round 3 | FS-Mol: A Few-Shot Learning Dataset of Molecules | fs-mol | 2021 | [129] |
| 22 | Round 3 | Evaluating Out-of-Distribution Performance on Document Image Classifiers | eval_ood | 2022 | [67] |
| 23 | Round 3 | Dungeons and Data: A Large-Scale NetHack Dataset | dungeons | 2022 | [37] |
| 24 | Round 3 | VisAlign: Dataset for Measuring the Alignment between AI and Humans in Visual Perception | visalign | 2023 | [70] |
| 25 | Round 3 | American Stories: A Large-Scale Structured Text Dataset of Historical U.S. Newspapers | amerstories | 2023 | [18] |
| 26 | Round 4 | KeSpeech: An Open Source Speech Dataset of Mandarin and Its Eight Subdialects | kespeech | 2021 | [133] |
| 27 | Round 4 | Habitat-Matterport 3D Dataset (HM3D): 1000 Large-scale 3D Environments for Embodied AI | habitat | 2021 | [112] |
| 28 | Round 4 | FLAIR: a Country-Scale Land Cover Semantic Segmentation Dataset From Multi-Source Optical Imagery | flair | 2023 | [31] |
| 29 | Round 4 | MedSat: A Public Health Dataset for England Featuring Medical Prescriptions and Satellite Imagery | medsat | 2023 | [120] |
| 30 | Round 4 | PUG: Photorealistic and Semantically Controllable Synthetic Data for Representation Learning | pug | 2023 | [9] |
| 31 | Round 5 | Constructing a Visual Dataset to Study the Effects of Spatial Apartheid in South Africa | spatial_apart | 2021 | [125] |
| 32 | Round 5 | A Spoken Language Dataset of Descriptions for Speech-Based Grounded Language Learning | spoken_lang | 2021 | [59] |
| 33 | Round 5 | Ambiguous Images With Human Judgments for Robust Visual Event Classification | ambiguous | 2022 | [118] |
| 34 | Round 5 | SCAMPS: Synthetics for Camera Measurement of Physiological Signals | scamps | 2022 | [91] |
| 35 | Round 5 | Objaverse-XL: A Universe of 10M+ 3D Objects | objaverse | 2023 | [17] |
| 36 | Round 5 | Timers and Such: A Practical Benchmark for Spoken Language Understanding with Numbers | timers | 2021 | [82] |
| 37 | Round 5 | CREAK: A Dataset for Commonsense Reasoning over Entity Knowledge | creak | 2021 | [102] |
| 38 | Round 5 | CLEVRER-Humans: Describing Physical and Causal Events the Human Way | clevrer | 2022 | [88] |
| 39 | Round 5 | OpenProteinSet: Training data for structural biology at scale | openprotein | 2023 | [2] |
| 40 | Round 5 | SSL4EO-L: Datasets and Foundation Models for Landsat Imagery | ssl | 2023 | [130] |
| 41 | Round 5 | OpenFilter: A Framework to Democratize Research Access to Social Media AR Filters | openfilter | 2022 | [114] |

| 42 | Round 5 | PROSPECT: Labeled Tandem Mass Spectrometry Dataset for Machine Learning in Proteomics | prospect | 2022 | [126] |
| 43 | Round 5 | MoCapAct: A Multi-Task Dataset for Simulated Humanoid Control | mocapact | 2022 | [136] |
| 44 | Round 5 | MADLAD-400: A Multilingual And Document-Level Large Audited Dataset | madlad | 2023 | [64] |
| 45 | Round 5 | Seeing is not always believing: Benchmarking Human and Model Perception of AI-Generated Images | seeing | 2023 | [76] |
| 46 | Round 5 | STAR: A Benchmark for Situated Reasoning in Real-World Videos | star | 2021 | [139] |
| 47 | Round 5 | BiToD: A Bilingual Multi-Domain Dataset For Task-Oriented Dialogue Modeling | bitod | 2021 | [75] |
| 48 | Round 5 | ActionSense: A Multimodal Dataset and Recording Framework for Human Activities Using Wearable Sensors in a Kitchen Environment | actionsense | 2022 | [19] |
| 49 | Round 5 | RenderMe-360: A Large Digital Asset Library and Benchmarks Towards High-fidelity Head Avatars | renderme | 2023 | [104] |
| 50 | Round 5 | PTADisc: A Cross-Course Dataset Supporting Personalized Learning in Cold-Start Scenarios | ptadisc | 2023 | [50] |
| 51 | Round 5 | ConfLab: A Data Collection Concept, Dataset, and Benchmark for Machine Analysis of Free-Standing Social Interactions in the Wild | conflab | 2022 | [113] |
| 52 | Round 5 | CSAW-M: An Ordinal Classification Dataset for Benchmarking Mammographic Masking of Cancer | csaw | 2021 | [127] |
| 53 | Round 5 | WikiChurches: A Fine-Grained Dataset of Architectural Styles with Real-World Challenges | wikichurches | 2021 | [4] |
| 54 | Round 5 | Mathematical Capabilities of ChatGPT | mathgpt | 2023 | [29] |
| 55 | Round 5 | SubseasonalClimateUSA: A Dataset for Subseasonal Forecasting and Benchmarking | subseas | 2023 | [97] |
| 56 | Round 5 | COVID-19 Sounds: A Large-Scale Audio Dataset for Digital Respiratory Screening | covid | 2021 | [140] |
| 57 | Round 5 | Shifts: A Dataset of Real Distributional Shift Across Multiple Large-Scale Tasks | shifts | 2021 | [86] |
| 58 | Round 5 | Wukong: A 100 Million Large-scale Chinese Cross-modal Pre-training Benchmark | wukong | 2022 | [36] |
| 59 | Round 5 | Addressing Resource Scarcity across Sign Languages with Multilingual Pretraining and Unified-Vocabulary Datasets | sign_lang | 2022 | [100] |
| 60 | Round 5 | SynMob: Creating High-Fidelity Synthetic GPS Trajectory Dataset for Urban Mobility Analysis | synmob | 2023 | [144] |

## D.2 Method for selecting datasets

In order to select datasets to evaluate, we first filtered all publications from the NeurIPS Datasets and Benchmarks track based on the following inclusion and exclusion criteria:

Inclusion:
1. Submitted paper creates a dataset.

Exclusion:
1. The paper discusses supplementary material and documentation, but it is not available.
2. The paper contributes a dataset but there are other contributions also made, which are discussed to a greater length, and the discussion of the dataset creation is less than one section of the paper.

Edge cases were discussed among the authors and categorized as included or excluded. From the remaining datasets, we randomly selected datasets to evaluate for each round.

### D.3 Evaluation consistency

**Additional note on methods:** In the final round, four reviewers performed double coding for 30 datasets, each reviewing on average 15 datasets. Accordingly, we measured IRR with a one-way mixed, consistent, average-measures intra-class coefficient (ICC). The final round, as with Round 3 and 4, also consisted of a disagreement review. After the disagreement review, additional corrections were made for consistency. This included:

1. If a reviewer changed their score from 'pass' to 'fail' for the minimum standard, the standard of excellence score was automatically changed to 'none'.
2. If a reviewer changed their score from 'fail' to `pass' for the minimum standard, the standard of excellence score was automatically changed to match the 2nd reviewer's score.

**Additional results (IRR):** We measured IRR per datasets and rounds as well as rubric categories. Specifically, the lowest ICC value for a given dataset in the initial rounds (training to round 4) was 0.45, indicating fair agreement, while the highest was 0.94, signifying excellent agreement. Subsequently, in the final round, the median ICC value for the 30 datasets evaluated was 0.90, with the highest ICC value for a given dataset as 1 indicating perfect agreement (Figure 4).

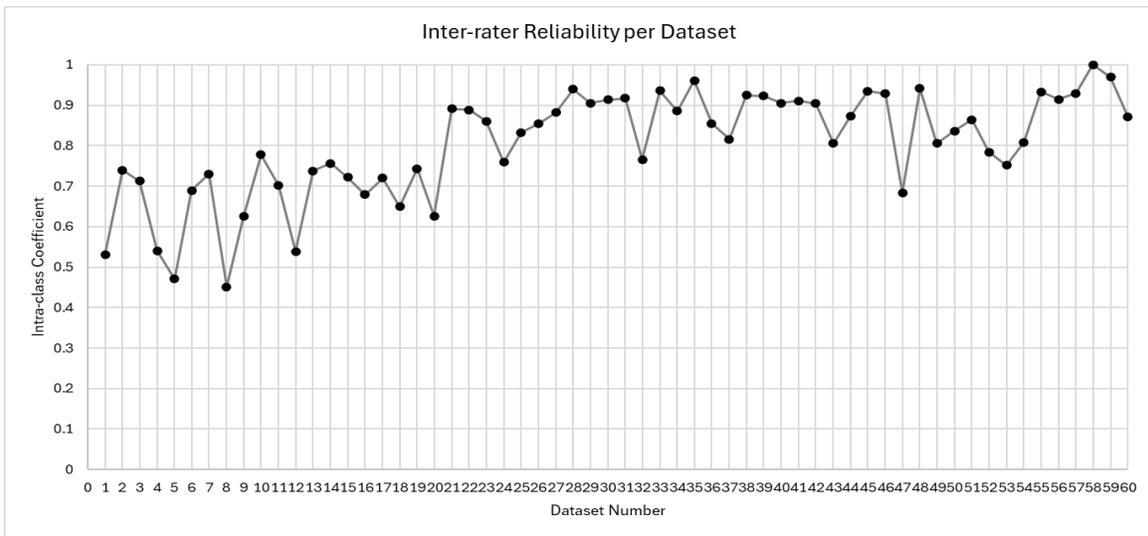

Figure 4. IRR for datasets. Datasets 1-30 measured with two-way ICC and 31-60 with one-way ICC.

Despite the greater level of interpretation present when evaluating IRR across elements as compared to datasets, the final round had a median ICC value >0.82 across all rubric categories (Figures 3a and 3b). Greater degree of variability can also be seen for the earlier rounds as compared to round 5, with round 5 ICC values having a median of 0.83-0.98 across rubric categories. Furthermore, many elements had perfect agreement especially for the minimum standard criteria ('context, purpose, motivation', 'context awareness', 'environmental footprint', 'data collection', 'data processing', 'data annotation', 'suitability', 'reliability', 'structured documentation', 'findability', and 'reusability'). The IRR for rubric categories from rounds 1-4 is shown in Figure 5.

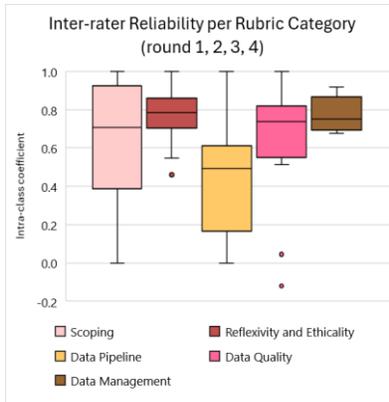

Figure 5. IRR for rubric categories calculated using two-way ICC for rounds 1-4.

**Additional results (consistency):** In addition to IRR, we measured inconsistency across evaluations by calculating the number of disagreements between reviewers. Disagreements were categorized in the following manner for training and rounds 1-4:

- Minor disagreement, standard of excellence (Minor, exc): instances where 1 of 3 reviewers disagree, e.g., Full, None, Full
- Major disagreement, minimum standard (Major, min): instances where 1 of 3 reviewers disagree, e.g., Pass, Pass, Fail
- Major disagreement, standard of excellence (Major, exc): instances all reviewers disagree, e.g., Full, Partial, None
- No disagreement, standard of excellence: instances where 1 of 3 reviewers gives a Partial evaluation and the other 2 reviewers agree, e.g., Partial, None, None

This was further simplified in round 5:

- Major disagreement, minimum standard (Major, min): instances where 1 of 2 reviewers disagree, e.g., Pass, Fail
- Major disagreement, standard of excellence (Major, exc): instances 1 of 2 reviewers disagree, e.g., Full, None
- No disagreement, standard of excellence: instances where 1 of 2 reviewers gives a Partial evaluation, e.g., Partial, None, None

We observed that the inconsistencies across datasets had markedly decreased by the final round. Figure 6 illustrates the trend in inconsistencies across datasets throughout multiple rounds of evaluations. The trend in major disagreements showed an even more pronounced reduction: for major inconsistencies under the minimum standard (Major, min), the inconsistency rates decreased significantly from initially high levels in the training and early rounds to much lower levels by Round 5. Similarly, major inconsistencies under the standard of excellence (Major, exc) see a sharp reduction, highlighting the impact of targeted improvements in rubric clarity and rater understanding. When considering all types of inconsistencies combined, the graph shows a substantial decrease from over 30% in the earliest phases to around 2% by the end of the final round, demonstrating nearly complete alignment among evaluators. This uniformity is indicative of the rubric's maturity as a tool for assessing dataset documentation quality.

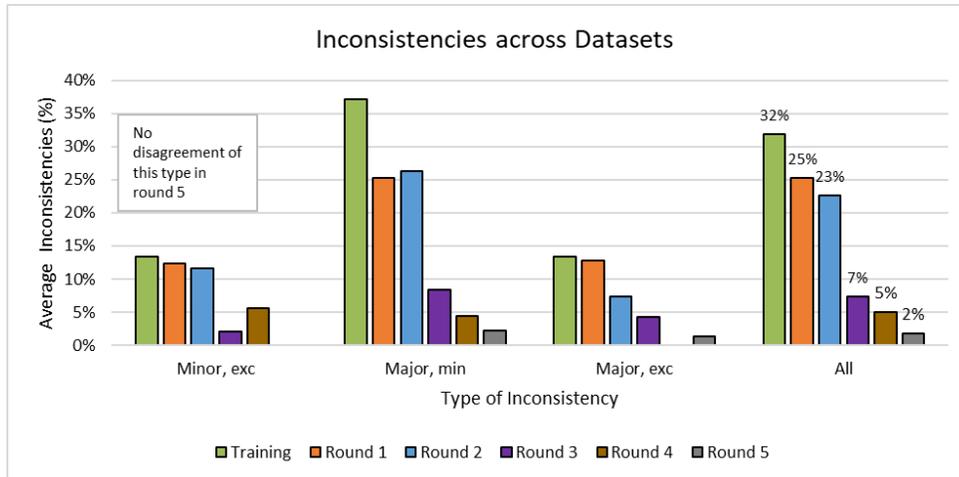

Figure 6. Inconsistencies in the rubric application across datasets.

In analyzing the inconsistencies among elements across several rounds, we found a clear trend of decreasing inconsistencies across almost all rubric elements. Progressing through rounds, the inconsistencies in even the most challenging elements had been markedly reduced (Figure 7a). By round 5, the percentage of inconsistencies reduced to 10% and under and only persisted for 7 of the 18 rubric elements (Figure 7b).

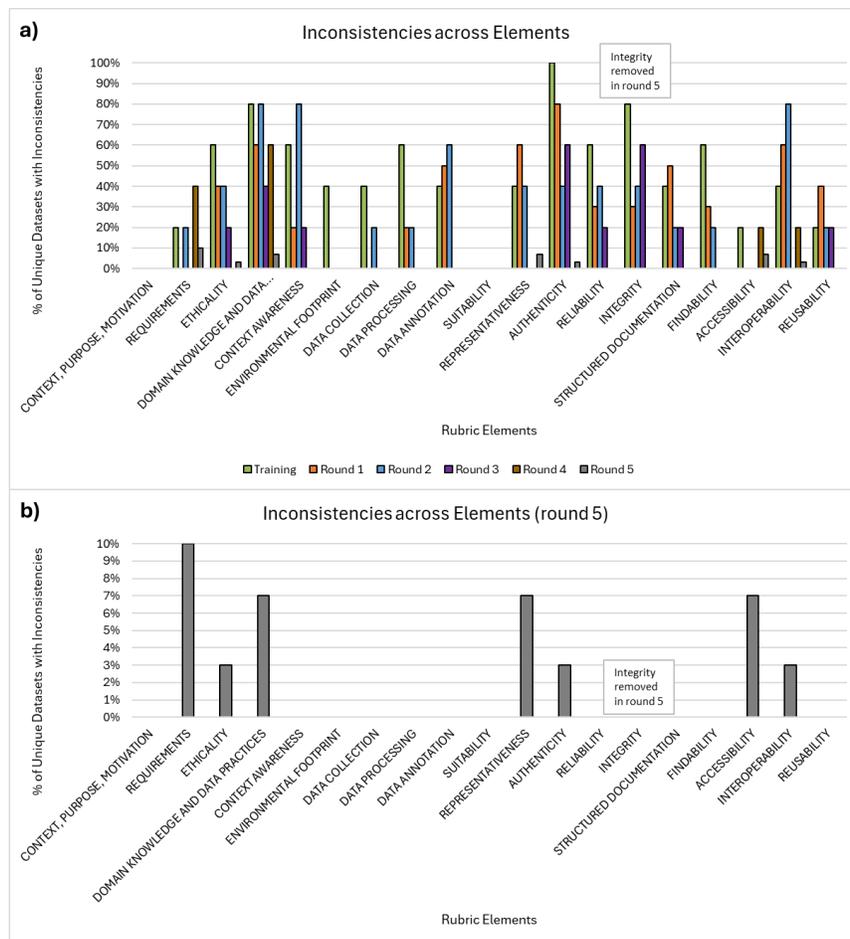

Figure 7. Inconsistencies in the rubric application. (a) Shows inconsistencies across elements, and particularly in the final round (b).

### D.4 Positionality, reflection, and contributions

*Positionality* begins by recognizing that knowledge is always produced from a certain perspective and that this perspective is always dependent on where we stand. Feminist standpoint theory emphasizes that there is no knowledge without a subject who knows. By identifying the location and situatedness of the knowing subject, we gain a better understanding of the knowledge in question. This makes the knowledge more robust, not less - it is a *stronger* version of objectivity than the flawed idea that knowledge can be fully disassociated from a subject and would embody a perspective of 'nowhere' [38–40]. By clarifying from where we speak, we can be more objective [94]. This is true for individuals as much as for groups. A key emphasis is often on recognized dimensions of intersectionality including but not limited to gender, age, sex, race, class, religion, dis/abilities, or geographies. In addition, the knowledges each team member brings to a group are also highly relevant.

Positionality statements can take many different shapes dependent on context and purpose. Beyond simply declaring aspects of identity, there is most value in *reflecting* how specifically these aspects *intersect* and shape the work being presented. Below, we will include positionality statements from each person disclosing what they are comfortable to disclose, followed by joint reflections on how the interaction of these perspectives have influenced this project.

Tegan: My primary academic training is in machine learning, specifically deep learning. Much of my master's and doctoral work examined generalization and learning behaviour of deep networks on real-world data (e.g. for climate and video data) using empirical methodology, giving me a strong and grounded appreciation for the importance of data in machine learning. As a professor, I am increasingly engaged in the interdisciplinary research I believe is necessary to make the field of machine learning/AI more responsible and empirically rigorous. All of my training and much of my perspective remain influenced by the dominant paradigms of the field, which generally have strong western, white, colonial/extractive, heteronormative, and patriarchal biases; my identity as a mixed-race, female-presenting, first-generation scholar has likely sensitized me to these influences. My work seeking change in the field (e.g. to our notions of novelty, representation, intelligence, rigor, or appropriate scientific practice) has been mostly internally reform-oriented.

Christoph: I am a white central European immigrant settler in Toronto, Canada, with an invisible disability. This identity has afforded me an odd juxtaposition of experiencing how privilege and oppression can intersect in our societies and has sensitized me over the years to theories and frameworks that help us make sense of these issues, including the intersectional feminist theories that shape this positionality statement. My education was in computer science and business informatics, but I am now a professor of information and am interacting closely with fields outside of computer science. My first published paper applied self-organizing maps to software cost estimation, but my doctoral and postdoctoral work developed computing and decision analysis frameworks for large-scale digital curation problems in libraries and archives. I became interested in long-term perspectives of environmental sustainability and the broader implications of technology on social justice partially because of the longer-term perspectives offered by ideals of stewardship and archiving. In the past decade I have grappled with the myths and illusions that are common in computing and that I too inherited via my education - including the flawed idea that technology is or can be neutral and that people's minds are information processors - and have built 'critical friendships' with other fields that can help computing researchers, educators and practitioners see beyond these conventional horizons and make better sense of the role of computing in our societies in order to help reorient computing for environmental sustainability and social justice. This negotiation between different fields and disciplines can be uncomfortable especially when it implies critique. In this project I aim to be pragmatic in building bridges between the conceptual spaces of these fields and to translate valuable insights from the stewardship perspective of digital curation into practical guidelines that ML can adopt and use.

Eshta: As a brown, female PhD student, I feel a lot of privilege in the opportunity to be a researcher where so many like me have not had the same opportunities, spaces, and circumstances. While I am a minority in these social identities, I am not in others and that is likely what has led to a position of privilege that I acknowledge. I am therefore also in a position where I can discuss and reflect on the impacts of my identities and worldviews on how and what I research. My academic background is rooted in information technology and data science, but my doctoral work is within a sociotechnical space. My perspective has thus shifted from seeing technology as a neutral entity, a tool that must be used and advanced further, to recognition of the need for a critical approach to technology that takes an ethics and justice based lens. The exposure to different worldviews that I have gained during my doctoral research has altered how I perceive the use of technology. For this project, I have pursued balancing a technical and

critical perspective so that the adoption of data curation that we recommend has practical benefits and uptake within the ML community.

Harshit: HG brings a perspective informed by his background in data science and information technology, shaped by his studies in South Asia. He is attuned to both the potential and the challenges of scaling technology in contexts vulnerable to climate change. In his work at the Department of Computer Science, HG evaluates how technological advancements can be balanced against their environmental impacts, striving to produce evidence that supports stringent environmental standards through a public health lens. His position is situated at the intersections between machine learning, environmental policy, and health outcomes.

Reyna: I am an Asian female PhD student. My primary academic training is in computer science and statistics, with a focus on climate informatics. Much of my academic research uses empirical methodologies to build machine learning and statistical models on real-world datasets, such as climate and scRNA-seq data. Throughout my PhD journey, I have recognized the importance of adopting a critical approach that incorporates ethics, responsibility, and sustainability in technology. This understanding is driven by my passion for studying climate change and the potential and limitations of machine learning to address it. I am particularly committed to promoting explainable AI, ethical AI, and responsible AI for climate change. Through my doctoral research, I have embraced a multifaceted and inclusive approach, striving to balance technical excellence with ethical considerations. This includes advocating for the responsible use of machine learning to tackle climate change emphasizing the importance of sustainability and social justice in my work.

Ciara: I am a PhD candidate at a faculty of information, where my primary research focus is on how people work with cross-domain data. I also work as a data scientist with the federal government, where I develop AI/ML governance and support the implementation of AI/ML projects. In my academic research, I draw on data and information practice scholarship, intersectional feminist and queer data studies, and interdisciplinary studies. My commitments to interdisciplinarity and feminist perspectives are influenced by my positions as a mixed-race woman, first generation university student, and settler in Canada. These commitments mean that I approached this project centering the situatedness of the rubric (that is: we are proposing a set of data curation best practices for ML rather than implying a sole universal 'right' way to do data work) and paying attention to the ways in which subject matter domain might have impacted the documentation of the datasets we evaluated.

Note: The fact that these statements above are all slightly different is an expression of their *authenticity* - we refrained from homogenizing them to fit a common structure, choosing instead to leave our individual voices here and present additional context.

Our perspectives interacted most directly in the making of the rubric. In our previous work, we discussed how disagreements in evaluations occurred because of differences in perspectives and areas of knowledge, including overlapping technologies, negotiating the depth of data curation expertise needed to apply the rubric, and challenges in scoping the extent of documentation dataset creators are responsible for [8]. These disagreements were deliberated to reach a balance between points of view. Particularly, it is the different points of view that enabled the rubric to take a shape in which both data curation and ML concepts were harmonized. When assembling the team, the senior researchers intentionally selected candidates with diverse perspectives. The composition of this team proved well-suited for doing the translation and operationalizing process of data curation for ML. Simply having two varying perspectives on the same dataset and the same criterion helped create a more nuanced view, and this triangulation of perspectives often enriched the debate and reflection on data practices. Other teams with diverse compositions with an interest in ML data practices as well as interdisciplinary complementarity may produce different assessment results, but a combination of perspectives will be valuable. We thus recommend having diversity and interdisciplinary complementarity in the application of the rubric.

We did not aim to neutralize the specifics of each individual viewpoint in a supposedly 'neutral' rubric - something that is never feasible - but instead sought *consistency* in the evaluation process. It is important to overcome misleading ideals of curation as neutral. As with other technical work, the neutral stance is illusory. Fields concerned with curation have also grappled with the realization that they cannot be neutral at all [111]. As a consequence, "current archival thought now recognizes and explores the implications of the subjective and inherently political nature of archival processes" such as appraisal: the decision what to keep and what to discard [122:162].

Our work in developing this framework extends a 'critical friendship' [14] from data curation towards machine learning. While machine learning already performs curation, it does so without adopting the field's standards which can aid in advancing the state of the art. Our aims were to clearly communicate knowledge from the data curation literature and communicate with machine learning researchers, in order to normatively encourage better practice.

Table 3: Author contributions

| Author | Contributions |
|---|---|
| Eshta Bhardwaj (she/her) | Her contributions for this project include the conceptualization of the framework, aiding with project administration, developing the methodology of conducting the evaluations, conducting evaluations, analyzing and visualizing the results, and writing, editing, and reviewing all drafts of the published work. |
| Harshit Gujral (he/him) | His contributions to this paper include the development and iteration of the evaluations, conducting evaluations, its writing, particularly methods, and results, and a discussion about reporting the environmental footprint of machine learning. |
| Siyi Wu (she/her) | Her contributions to this paper include development and iteration of evaluations, conducting evaluations, writing, review and editing. |
| Ciara Zogheib (she/her) | Her contributions include conducting iterative evaluations of dataset documentation, and writing (results section, positionality) and edits of the final manuscript. |
| Tegan Maharaj (they/them) | Their contributions to this paper include conceptualization, funding, methodology, comments on iterations of the rubric, writing, review and editing. |
| Christoph Becker (he/him) | His contributions include conceptualization, resources, funding, methodology, supervision, validation, writing, reviewing and editing. |

### D.5 How to report environmental footprint

We recommend the following strategies to quantify the environmental footprint of dataset development:

1. Carbon Footprint Estimation Tools: Tools like LLMCarbon, which has been designed to provide end-to-end carbon footprint estimations for large language models, offer a valuable resource for dataset creators in ML [78]. These tools allow for the prediction of carbon outputs based on diverse parameters such as hardware use, model architecture, and operational practices before the actual computational tasks begin.
2. Efficient Data Management Strategies: Research shows that end-to-end utilization of each GB stored in a data center is associated with approximately 5.12 kWh of energy consumption [16]. Reducing data redundancy and implementing data pruning techniques can decrease the volume of data that needs to be stored and processed, subsequently reducing the energy consumed during these stages. Several researchers have called for using end-to-end efficiency as an evaluation criterion for publishing ML research on computationally intensive models besides accuracy and related measures [43, 106, 124, 131].
3. Standardize Carbon Reporting: Developing a standardized protocol for reporting the carbon emissions of ML projects, as suggested by existing research [65, 78, 80, 106], would facilitate greater transparency and accountability within the industry. For the very least, researchers can gather a rough estimate of the electricity consumption of their ML projects and can get an estimate of corresponding carbon dioxide equivalent emissions using tools like the ML Emissions Calculator [65] and Green Algorithms tool [66]. This could also involve detailed reporting of energy sources, hardware specifications, and operational efficiencies.
4. Large but Sparsely Activated Networks: The previous research discusses the energy efficiency of using large but sparsely activated deep neural networks (DNNs), which can consume less than one-tenth the energy of large, densely activated DNNs without sacrificing accuracy [106]. For data processing, training, and deployment, the paper also emphasizes the impact of choosing energy-efficient data centers and hardware, along with strategically choosing data centers at a geographic location with a high renewable energy mix in the electricity grid [106].
5. Life cycle assessment (LCA) approach: For large language models conducting a Life Cycle Assessment (LCA) is recommended to quantify the carbon footprint across all stages of a language model's life cycle, from

equipment manufacturing to operational use and beyond [79]. This methodology, as discussed in the BLOOM model analysis [79], includes the energy consumed during model training and the emissions during model deployment and inference.

We hope these recommendations provide a much-needed starting point for the ML community to begin quantifying the environmental footprint of their projects. However, reporting alone will not mitigate the environmental impacts of data curation in ML. Existing research encourages the ML community to engage in efficient data management strategies, optimize model architectures, and adopt green computing practices like selecting data centers that use renewable energy [43, 106, 124, 131]. Research suggests that while efficiency improvements are crucial, they are not always sufficient on their own to reduce overall carbon emissions due to these rebound effects [73, 128].

Rebound effects manifest when improvements in computational efficiency lead to an increased use of ML technologies, as lower operational costs and enhanced capabilities encourage more frequent training and deployment of larger models, potentially increasing total energy consumption [12, 15, 128]. Historical data in sectors like automotive and residential energy use demonstrate that efficiency gains often lead to increased consumption, as savings are reinvested into more or expanded use of the technology, rather than resulting in a net decrease in energy use [12, 52, 134]. This can be exacerbated by indirect rebound effects that occur when the application of energy-efficient ML technologies in various sectors leads to broader and more intensive use of these technologies, subsequently increasing energy demand across those sectors, despite individual efficiency improvements.

In addition to energy efficiency-based measures, we call for the inclusion of digital sufficiency-based measures to mitigate potential rebound effects [73, 119]. These measures include limiting the growth of computational demands by setting strict computational budgets that reflect actual needs rather than maximum capacities. Additionally, it involves the creation of algorithms designed to perform effectively with minimal energy use, emphasizing necessity over excess. Regulatory measures are also critical, aimed at enforcing sustainable practices across the digital and computing sectors to ensure that efficiency improvements result in genuine reductions in carbon emissions. These mitigation strategies integrate sufficiency with technological innovation, ensuring that the advancement of ML contributes positively to environmental sustainability.

### E. Changes to Rubric and Toolkit

Table 4: List of Changes to Rubric and Toolkit

| Location of Change | Description and Rationale |
|---|---|
| Toolkit: Application Guidance | The evaluation of the minimum standard of documentation was updated. It previously stated that a pass is granted for *any* amount or type of discussion around the element and a fail is granted *only* if there is no discussion around the element at all. In the new version of the toolkit, it states that a pass is granted if all aspects specified under the minimum standard were discussed and a fail if they were only partially discussed or not discussed at all.<br><br>Based on the changes made to criteria of the rubric elements, the minimum standard could only be achieved if all criteria were fulfilled (not just one or few). |
| Toolkit: How to interpret authenticity, reliability, and representativeness | An additional example of how to interpret authenticity and reliability was added to aid in better understanding of the criteria of the elements. |
| Toolkit: Sample evaluations | Both sample evaluations were updated. Sample evaluations were updated to reflect the completion of a more recent version of the rubric. |
| Rubric: Context, purpose, motivation | The criteria for the standard of excellence were made more explicit so that evaluators would have similar interpretations. Previously it stated, "documentation explains how dataset can be reused beyond its original context". The current version expects documentation to discuss whether and how reuse is possible. |
| Rubric: Requirements | The criteria for the standard of excellence were simplified, and the requirement to "state different approaches in formulating the problem apart from the final presented plan" was removed. |

| | |
|---|---|
| | The criterion was simplified because we recognized that documenting alternative approaches was not an essential characteristic of excellence: there are good reasons not to document paths not chosen. |
| Rubric: Context awareness | The criterion for the minimum standard was simplified by moving the requirement for a "reflection on the dataset creators' awareness of social, political, and historical context" to the standard of excellence criteria for 'context, purpose, motivation'.<br><br>The criterion was moved for clarity: the reflection on context ties in with purpose and motivation, while the 'context awareness' dimension focuses on positionality and reflexivity. |
| Rubric: Domain knowledge and data practices | The phrasing for both criteria were updated. The phrasing conveys similar criteria as the original, but is more explicitly stated to enable more consistent evaluation. |
| Rubric: Data collection | The criteria for both minimum standard and standard of excellence were updated to include multiple types of data.<br><br>The criteria were augmented to address both collected data and synthetic data because it originally did not ask for documentation regarding how data was synthesized and whether that introduces intrinsic biases. |
| Rubric: Data annotation | The criteria for both minimum standard and standard of excellence were updated to include the assessment of labels obtained from multiple sources.<br><br>The previous criteria did not have differing criteria based on how the labels were obtained/derived. |
| Rubric: Authenticity | We combined the criteria for authenticity to include integrity based on [55] to simplify the rubric. |
| Rubric: Reliability | Based on feedback from the reviewers, we added a clarifying statement to make the criteria for the minimum standard clearer. We also changed the criteria for the standard of excellence to clarify how it can be evaluated. |
| Rubric: Structured documentation | We slightly updated the phrasing to only allow context documents to have established structures rather than formats that are not established and therefore may be difficult to evaluate consistently. |
| Rubric: Reusability | Minor changes to the phrasing were made to the minimum standard to improve clarity. |



\end{document}